\documentclass[11pt]{article} 

\usepackage[ansinew]{inputenc}
\usepackage{amsmath, amssymb, graphics, amsthm}
\usepackage[pdftex]{graphicx, color}
\usepackage{epsfig}
\usepackage{color}
\usepackage{undertilde}
\usepackage{fancyhdr} 

\usepackage{hyperref}

\newtheorem*{theorem}{Theorem}

\newtheorem*{lemma}{Lemma}

\newcommand{\m}{\mbox{}}

\newcommand{\ba}{\begin{equation}}
\newcommand{\ea}{\end{equation}}
\newcommand{\be}{\begin{eqnarray}}
\newcommand{\ee}{\end{eqnarray}}

\makeatletter
\@addtoreset{equation}{section}
\makeatother

\title{{\sf New Variables for Classical and Quantum Gravity}\\
{\sf in all Dimensions III. Quantum Theory}} 
\author{
{\sf N. Bodendorfer}$^{1,2}$\thanks{{\sf 
norbert.bodendorfer@gravity.fau.de}},
{\sf T. Thiemann}$^{1,3}$\thanks{{\sf 
thomas.thiemann@gravity.fau.de,
tthiemann@perimeterinstitute.ca}},
{\sf A. Thurn}$^1$\thanks{{\sf 
andreas.thurn@gravity.fau.de}}\\
\\
{\sf $^1$ Inst. for Theoretical Physics III, FAU Erlangen -- N\"urnberg,}\\
{\sf Staudtstr. 7, 91058 Erlangen, Germany}\\
\\
{\sf $^2$ Institute for
Gravitation and the Cosmos \& Physics
  Department,}\\
{\sf   Penn State, University Park, PA 16802, U.S.A.}\\
\\
{\sf $^3$ Perimeter Institute for Theoretical Physics,}\\ 
{\sf 31 Caroline Street N, Waterloo, ON N2L 2Y5, Canada}
}
\date{{\small\sf \today}}

\begin{document} 

\maketitle

\normalsize

{\sf

\begin{abstract}
We quantise the new connection formulation of $D+1$ dimensional General Relativity developed in our
companion papers by Loop Quantum Gravity (LQG) methods. It turns out that all the tools
prepared for LQG straightforwardly generalise to the new connection formulation in higher
dimensions. The only new challenge is the simplicity constraint. While its ``diagonal'' components
acting at edges of spin network functions are easily solved, its ``off-diagonal'' components acting at vertices are non trivial and require a more elaborate treatment.    
\end{abstract}

}

\newpage
\tableofcontents
\newpage

\section{Introduction}

In our companion papers \cite{BTTI,BTTII} we developed the classical framework for a new connection
formulation of General Relativity that is applicable in all spacetime dimensions $D+1\ge 3$. 
In $3+1$ dimensions, the current connection formulation is based on a triad and its corresponding
spin connection. The miracle that happens in three spatial dimensions is that the defining
representation of SO$(3)$ is equivalent to its adjoint representation. Therefore, a connection
and a triad carry the same number of degrees of freedom and can serve as a canonical
pair on an extended phase space whose reduction by the SO$(3)$ Gau{\ss} constraint leads back
to the ADM phase space. In order that the connection is Poisson commuting, a further miracle
has to happen, namely the spin connection is integrable, i.e. can be obtained from a functional
by functional derivation.  These two miracles are reserved for $D=3$. The observation that 
enables a connection formulation in higher dimensions as well is that the mismatch between
the number of degrees of freedom of the $D$-bein and its spin connection can be accounted for by
a new constraint in addition to the Gau{\ss} constraint, which requires that the momentum conjugate
to the connection comes from a $D$-bein. The details are a bit more complicated, we have 
to use SO$(D+1)$ rather than SO$(D)$, the $D$-bein is a generalised $D$-bein and the spin connection 
is a generalised hybrid connection,  but this is the rough idea.

The final picture is therefore a SO$(D+1)$ gauge theory subject to SO$(D+1)$ Gau{\ss} constraint,
simplicity constraint, spatial diffeomorphism constraint and Hamiltonian constraint. Apart from 
the different gauge group which however is compact and the additional simplicity constraint,
the situation is precisely the same as for LQG and the quantisation of our connection formulation
is therefore in complete analogy with LQG. We can therefore simply follow any standard 
text on LQG such as \cite{RovelliQuantumGravity, ThiemannModernCanonicalQuantum} and follow all the quantisation steps. This way 
we arrive at the holonomy-flux algebra, its unique spatially diffeomorphism invariant 
state whose GNS data are the analogue for SO$(D+1)$ of the Ashtekar-Isham-Lewandowski Hilbert space, the analogue of spin network functions, kinematical geometrical operators such as the volume operator which is pivotal
for the quantisation of the Hamiltonian constraint, the SO$(D+1)$ Gau{\ss} constraint, the spatial
diffeomorphism constraint, the Hamiltonian constraint and a corresponding Master constraint.

The only structurally new ingredient is the simplicity constraint which constrains the type 
of allowed SO$(D+1)$ representations. When it acts at the interior point of edges, it 
requires that the corresponding SO$(D+1)$ representation is simple. However, when it
acts at a vertex, the constraint splits into several linearly independent ones which 
are not mutually commuting and do not close on themselves. The situation here is 
similar to the situation in spin foam models \cite{EngleTheLoopQuantum, LivineNewSpinfoamVertex, EngleFlippedSpinfoamVertex, EngleLoopQuantumGravity, FreidelANewSpin, KaminskiSpinFoamsFor} where similar constraints at 
the discretised level for SO$(4)$ arise while ours are for SO$(D+1)$ in the continuum. We 
propose to solve these anomalous components of the simplicity constraints as in \cite{EngleTheLoopQuantum, LivineNewSpinfoamVertex, EngleFlippedSpinfoamVertex, EngleLoopQuantumGravity}  
by passing 
to a corresponding Master constraint and subtracting its spectral gap\footnote{The fact 
that the gauge group is compact makes sure that the spectrum of this Master constraint
is pure point.}.\\
\\
The manuscript is organised as follows:\\
\\
In section two we define the SO$(D+1)$ holonomy-flux algebra and the corresponding 
Hilbert space representation. In section three we implement the kinematical constraints,
that is Gau{\ss}, simplicity and spatial diffeomorphism constraints. In section four we develop
kinematical geometrical operators, specifically $D$-dimensional area and volume operators.
Lower dimensional operators such as length operators etc. can be constructed similarly
but are left for future publication. Finally, in section five we quantise the Hamiltonian constraint.
The presentation will be brief since all the constructions literally parallel those of LQG.
We therefore refer the interested reader to \cite{ThiemannModernCanonicalQuantum} for all the missing details.

\normalsize

\section{Kinematical Hilbert Space}
 
The construction of the kinematical Hilbert has been performed in \cite{AshtekarRepresentationsOfThe, AshtekarRepresentationTheoryOf, AshtekarDifferentialGeometryOn, AshtekarProjectiveTechniquesAnd, MarolfOnTheSupport, AshtekarQuantizationOfDiffeomorphism} for four and higher space-time dimension and arbitrary compact gauge group. These results apply for the case considered here, since we are using the compact group SO$(D+1)$ irrespective of the signature of the space-time metric. We therefore only cite the main results in this section and introduce notation needed later on. 

Since the Poisson brackets between $A_{aIJ}$ and $\pi^{bKL}$ are singular, we have to smear them with test functions. In order to obtain non-distributional Poisson brackets, smearing has to be done at least $D$-dimensional in total. $A_{aIJ}$ is a one-form, thus naturally smeared along a one-dimensional curve. From $\pi^{aIJ}$, being a vector density of weight one, we can construct the so$(D+1)$ - valued pseudo $(D-1)$-form $(* \pi)_{a_1 ... a_{D-1}} := \pi^{aIJ} \epsilon_{a a_1 ... a_{D-1}} \tau_{IJ}$ which is integrated over a $(D-1)$-dimensional surface in a background-independent way. These considerations lead to the definitions of holonomies and fluxes, which yield a natural starting point for a background independent quantisation. In the following, we choose $\left(\tau_{IJ}\right)^K\m_L = \frac{1}{2} \left( \delta^K_{I}\delta_{JL} - \delta^K_{J}\delta_{IL} \right)$ as a basis of the Lie algebra so$(D+1)$.

\subsection{Holonomies, Distributional Connections, Cylindrical Functions, Kinematical Hilbert Space and Spin-Network States}
\label{sec:Kin1}
Denote by $\mathcal{A}$ the space of smooth connections over $\sigma$. We define the holonomy $h_c(A) \in \text{SO}(D+1)$ of the connection $A \in \mathcal{A}$ along a curve $c:[0,1]\rightarrow \sigma$ as the unique solution to the differential equation
\be \frac{d}{ds} h_{c_s} (A) = h_{c_s} (A) A(c(s)), ~~ h_{c_0} = 1_{D+1}, ~~ h_c(A) = h_{c_1} (A) \text{,} \ee
where $c_s(t) := c(st)$, $s \in [0,1]$, $A(c(s)) := A_a^{IJ} (c(s)) \tau_{IJ} \dot{c}^a(s)$. The solution is explicitly given by
\be h_c(A) = \mathcal{P} \exp \left( \int_c A \right) = 1_{D+1} + \sum_{n=1}^\infty \int_0^1 dt_1 \, \int_{t_1}^1 dt_2 \ldots \int_{t_{n-1}}^1 dt_n A(c(t_1)) \ldots A(c(t_n)) \text{,} \ee
where $\mathcal{P}$ denotes the path ordering symbol which orders the smallest path parameter to the left. Like in $3+1$ dimensional LQG, we will restrict ourselves to piecewise analytic and compactly supported curves.

The holonomies coordinatise the classical configuration space. In quantum field theory it is generic that the measure underlying the scalar product of the theory is supported on a distributional extension of the classical configuration space. For gravity, this enlargement of the configuration space is done by generalising the idea of a holonomy. Since the equations
\be h_{c \circ c'}(A) = h_c(A) h_{c'}(A) \hspace{5mm} h_{c^{-1}}(A) = h_{c}(A)^{-1} \label{eq:hol}\ee
hold, we see that an element $A \in \mathcal{A}$ is a homomorphism from the set of piecewise analytic paths with compact support $\mathcal{P}$ into the gauge group. We now introduce the set \mbox{$\overline{\mathcal{A}}:=\text{Hom}(\mathcal{P}, \text{SO}(D+1))$} of all algebraic homomorphisms (without continuity assumptions) from $\mathcal{P}$ into the gauge group. This space $\overline{\mathcal{A}}$ is called the space of distributional connections over $\sigma$ and constitutes the quantum configuration space. The algebra of cylindrical functions $\text{Cyl}(\overline{\mathcal{A}})$ on the space of distributional SO$(D+1)$ connections is chosen as the algebra of kinematical observables. The former algebra can be written as the union of the set of functions of distributional connections defined on piecewise analytic graphs $\gamma$, $\text{Cyl}(\overline{\mathcal{A}}) = \cup_{\gamma} \text{Cyl}_{\gamma}(\overline{\mathcal{A}})/\sim$. $\text{Cyl}_{\gamma}(\overline{\mathcal{A}})$ is defined as follows. A piecewise analytic graph $\gamma \in \sigma$ consists of analytic edges $e_1$,...,$e_n$, which meet at most at their endpoints, and vertices $v_1$,...,$v_m$. We denote the edge and vertex set of $\gamma$ by $E(\gamma)$ ($|E(\gamma)| = n$) and $V(\gamma)$ ($|V(\gamma)| = m$), respectively. A function $f_{\gamma} \in \text{Cyl}_{\gamma}(\overline{\mathcal{A}})$ is labelled by the graph $\gamma$ and typically looks like $f_{\gamma}(A) = F_{\gamma}\left(h_{e_1}(A),...,h_{e_{|E|}}(A)\right)$, where $F_{\gamma}: \text{SO}(D+1)^{|E|} \rightarrow \mathbb{C}$. One and the same cylindrical function $f \in \text{Cyl}(\overline{\mathcal{A}})$ can be represented on different graphs leading to cylindrically equivalent representations of that function. It is understood in the above union that such functions are identified. We will denote the pullback of a function $f_{\gamma}$ defined on $\gamma$ on the bigger\footnote{The graph $\gamma$ can be enlarged by e.g. adding or subdividing edges. See e.g. \cite{ThiemannModernCanonicalQuantum} for a precise definition of the partial order on tame subgroupoids defined by graphs.} graph $\gamma' \succ \gamma$ via the cylindrical projections by $p^*_{\gamma'\gamma}$. Then, the equivalence relation just mentioned can be made more explicit, $f_{\gamma} \sim f'_{\gamma'}$ iff $p^*_{\gamma''\gamma}f_{\gamma} = p^*_{\gamma''\gamma'}f'_{\gamma'} \hspace{1mm} \forall \gamma,\gamma' \prec \gamma''$. The pullback on the projective limit function space will be denoted by $p^*_{\gamma}$. The functions cylindrical with respect to a graph that are $N$ times differentiable with respect to the standard differentiable structure on SO$(D+1)$ will be denoted by $\text{Cyl}^N_{\gamma}(\overline{\mathcal{A}})$ and $\text{Cyl}^N(\overline{\mathcal{A}}) := \cup_{\gamma} \text{Cyl}^N_{\gamma}(\overline{\mathcal{A}})/\sim$.

Since in the end we are interested only in gauge invariant quantities, after solving the Gau{\ss} constraint (classically oder quantum mechanically) we have to consider the algebra of cylindrical functions on the space of distributional connections modulo gauge transformations $\text{Cyl}(\overline{\mathcal{A}/\mathcal{G}})$. For representatives $f_{\gamma}$ of elements $f$ of this space, the complex-valued function $F_{\gamma}$ on $\m{\text{SO}(D+1)^{|E|}}$ has to be such that $f_{\gamma}(A)$ is gauge invariant. We will slightly abuse notation and use the same notation for the new projectors $p_{\gamma' \gamma} : \mathcal{A}_{\gamma'}/\mathcal{G}_{\gamma'} \rightarrow \mathcal{A}_{\gamma}/\mathcal{G}_{\gamma}$. There is a unique \cite{LewandowskiUniquenessOfDiffeomorphism, FleischhackRepresentationsOfThe} choice of a diffeomorphism invariant, faithful measure $\mu_0$ on $\overline{\mathcal{A}/\mathcal{G}}$ which equips us with a kinematical, gauge invariant Hilbert space $\mathcal{H}^0 := L_2\left(\overline{\mathcal{A}/\mathcal{G}}, d\mu_0\right)$ appropriate for a representation in which $A$ is diagonal. This measure is entirely characterised by its cylindrical projections defined by
\be\int_{\overline{\mathcal{A}/\mathcal{G}}} d\mu_0(A)f(A) &=& \int_{\overline{\mathcal{A}/\mathcal{G}}} d\mu_{0,\gamma}(A) f_{\gamma}\left(A\right) \nonumber \\
&=& \int_{\text{SO}(D+1)^{|E(\gamma)|}} \left[\prod_{e \in E(\gamma)} d\mu_H(h_e)\right] ~ F_{\gamma}\left(h_1,...,h_{|E|}\right)\text{,}\ee
where $\mu_H$ is the Haar probability measure on SO$(D+1)$. 

An orthonormal basis on $\mathcal{H}^0$ is given by spin-network states \cite{RovelliSpinNetworksAnd, BaezSpinNetworksIn, ThiemannTheInverseLoop}, which are defined as follows. Given a graph $\gamma$, label its edges $e \in E(\gamma)$ with non-trivial irreducible representations $\pi_{\Lambda_e}$ of SO$(D+1)$, i.e. $\Lambda_e$ is the highest weight vector associated with $e$, and its vertices $v \in V(\gamma)$ with intertwiners $c_v$, i.e. matrices which contract all the matrices $\pi_{\Lambda_e}(h_e)$ for $e$ incident at $v$ in a gauge invariant way. A spin-network state is simply a $C^{\infty}$ cylindrical function on $\overline{\mathcal{A}/\mathcal{G}}$ constructed on the above defined so-called spin-net, $T_{\gamma,\vec{\Lambda},\vec{c}}[A] := \text{tr} \left[\otimes_{i=1}^{|E|} \pi_{\Lambda_{e_i}}(h_{e_i}(A)) \cdot \otimes_{j=1}^{|V|} c_j\right]$, where $\vec{\Lambda} = (\Lambda_e)$, $\vec{c} = (c_v)$ have indices corresponding to the edges and vertices of $\gamma$ respectively.

\subsection{(Electric) Fluxes and Flux Vector Fields}
Since $\pi^{aIJ}$ are Lie algebra-valued vector densities of weight one, $(* \pi)_{a_1 ... a_{D-1}} := \pi^{aIJ} \epsilon_{a a_1 ... a_{D-1}} \tau_{IJ}$ is a pseudo $(D-1)$-form and is naturally integrated over a $(D-1)$-dimensional face $S$. We therefore define the (electric) fluxes 
\be \pi^n(S) := \int_S n_{IJ} (* \pi)^{IJ} = \int_S n_{IJ} \pi^{aIJ} \epsilon_{a a_1 \ldots a_{D-1}} dx^{a_1} \wedge \ldots \wedge dx^{a_{D-1}} \text{,} \label{eq:fluxes} \ee
where $n = n^{IJ} \tau_{IJ}$ denotes a Lie algebra-valued scalar function of compact support. We again restrict to piecewise analytic surfaces $S$, to ensure finiteness of the number of isolated intersection points of $S$ with a piecewise analytic path. In order to compute Poisson brackets, we have to suitably regularise the holonomies and fluxes to objects smeared in $D$ spatial dimensions. A possible regularisation in any dimension is given in \cite{ThiemannModernCanonicalQuantum}. Removal of the regulator leads to the following action of the Hamiltonian vector fields $Y_n(S)$ corresponding to $\pi_n(S)$ on adapted representatives $f_{\gamma_S}$
\be Y^n_{\gamma_S}(S)\left[f_{\gamma_S}\right] &=& \sum_{e \in E(\gamma_S)} \epsilon(e,S) \hspace{1mm} \left[n(b(e)) \hspace{1mm} h_e(A)\right]_{AB} \hspace{1mm} \frac{\partial F_{\gamma_S}}{\partial h_e(A)_{AB}} \left( h_{e_1}(A),...,h_{e_{|E(\gamma_S)|}(A)} \right) \nonumber \\ 
&=& \sum_{e \in E(\gamma_S)} \epsilon(e,S) \hspace{1mm} n^{IJ}(e \cap S) \hspace{1mm} R_{IJ}^e f_{\gamma_S} \label{eq:FluxAction} \text{.}\ee
$f_{\gamma_S}$ is an adapted representative of the cylindrical function $f \in Cyl^1(\overline{\mathcal{A}})$ in the sense that all intersection points of $S$ and $\gamma_S$ are beginning points $b(e)$ of edges $e \in E(\gamma_S)$ (this can always be achieved by suitably splitting and inverting edges). In the above equation, $\epsilon(e,S)$ is a type-indicator function, which is $+(-)1$ if the beginning segment of the edge $e$ lies above (below) the surface $S$ and zero otherwise. $R_{IJ}^{e}$ ($L^e_{IJ}$) is the right (left) invariant vector field on the copy of SO$(D+1)$ labelled by $e$,
\be \left(R_{IJ}f\right)(h) := \left(\frac{d}{dt}\right)_{t=0} f(e^{t \tau_{IJ}}h) \hspace{4mm} \text{and}	\hspace{4mm}	\left(L_{IJ}f\right)(h) := \left(\frac{d}{dt}\right)_{t=0} f(he^{t \tau_{IJ}}) \text{.} \ee
The algebra of right (left) invariant vector fields is given by
\be \left[R^e_{IJ},R^{e'}_{KL}\right] &=&  \frac{1}{2} \delta_{e,e'} \left(\eta_{JK} R^e_{IL} + \eta_{IL} R^e_{JK} -\eta_{IK} R^e_{JL}-\eta_{JL} R^e_{IK}\right)\text{,} \nonumber \\
  \left[R^e_{IJ},L^{e'}_{KL}\right] &=& 0 \text{,} \label{eq:LieAlgebraRelations}\ee
and analogously for $L^e_{IJ}$. We remark that, in order to calculate functional derivatives, we had to restrict $f$ to $\mathcal{A}$ in the beginning. The end result (\ref{eq:FluxAction}), however, can be extended to all of $\overline{\mathcal{A}}$. Following the standard treatment, these vector fields are generalised from adapted to non-adapted graphs and shown to yield a cylindrically consistent family of vector fields, thus they define a vector field $Y_n(S)$ on $\overline{\mathcal{A}}$. The $Y_n(S)$ are called flux vector fields. 

On the Hilbert space defined in section \ref{sec:Kin1}, the elements of the classical holonomy-flux algebra become operators which act by
\be 
	\hat{f} \cdot \psi &:=& f \hspace{1mm} \psi \text{,} \nonumber \\
	\hat{Y}_n(S) \cdot \psi &:=& i \hbar \kappa \beta Y_n(S)[\psi] \text{,}
\ee
where the right hand side is the action of the vector field $Y_n(S)$ on the cylindrical function $\psi$. The appearance of $\beta$ is due to the fact that we defined the fluxes using $\pi$, whereas the momenta conjugate to the connection is given by $\m^{(\beta)}\pi = \frac{1}{\beta} \pi$. The momentum operators $\hat{Y}_n(S)$, with dense domain $Cyl^1$, can be shown to be essentially self-adjoint operators on $\mathcal{H}^0$ analogously to the $(3+1)$-dimensional case \cite{AshtekarDifferentialGeometryOn}.

\section{Implementation and Solution of the Kinematical Constraints}

\subsection{Gau{\ss} Constraint}
Working with the gauge invariant Hilbert space from the beginning, the Gau{\ss} constraint is already solved. Yet we want to summarise its implementation on the gauge variant Hilbert space $\mathcal{H} = L_2\left(\overline{\mathcal{A}},d\mu'_0 \right)$, since we want to compute quantum commutators of the constraint with the simplicity constraint in the next section. The implementation (as well as the solution) of the Gau{\ss} constraint can be copied from the $(3+1)$-dimensional case without modification. 

According to the RAQ programme, we choose the dense subspace $\Phi = Cyl^{\infty}(\overline{\mathcal{A}})$ in the Hilbert space. Then, we are looking for an algebraic distribution $L \in \Phi'$ such that the following equation holds
\be L\left(p^*_{\gamma}\left[\sum_{e \in E(\gamma); \hspace{1mm} v= b(e)} R_{IJ}^e \hspace{2mm} - \sum_{e \in E(\gamma); \hspace{1mm} v= f(e)} L_{IJ}^e \right]f_{\gamma} \right) = 0 \ee
for any $v \in V(\gamma)$, any graph $\gamma$ and $f_{\gamma} \in Cyl_{{\gamma}}^{\infty}(\overline{\mathcal{A}})$. The general solution for $L$ is given by a linear combination of $\left\langle \psi,.\right \rangle$, where $\psi \in \mathcal{H}^0$ is gauge invariant. Thus, for an adapted graph $\gamma'$ (all edges outgoing from the vertex $v$ in question), gauge invariance amounts to vanishing sum of all right invariant vector fields at a vertex,
\be \sum_{e \in E(\gamma'); \hspace{1mm} v= b(e)} R_{IJ}^e f_{\gamma'} = 0 \text{.}\ee

\subsection{Simplicity Constraint}
\subsubsection{From Classical to Quantum}
\label{sec:Simplicity1}
Classically, vanishing of the simplicity constraints $S^{ab}_{\overline{M}}(x) = \frac{1}{4} \epsilon_{IJKL\overline{M}} \pi^{aIJ}(x) \pi^{bKL}(x)$ at all points $x \in \sigma$ is completely equivalent to the vanishing of
\be
	C_{\overline{M}}(S^x,S'^x) := \lim_{\epsilon,\epsilon' \rightarrow 0} \frac{1}{\epsilon^{(D-1)}\epsilon^{'(D-1)}} \epsilon_{IJKL\overline{M}} \pi^{IJ}(S^x_{\epsilon}) \pi^{KL}(S'^x_{\epsilon'})
\ee
for all points $x \in \sigma$ and all surfaces $S^x_{\epsilon}, S'^x_{\epsilon'} \subset \sigma$ containing $x$ and shrinking to $x$ as $\epsilon$, $\epsilon'$ tend to zero. More precisely, we use faces of the form $S^x: \hspace{1mm} (-1/2,1/2)^{D-1} \rightarrow \sigma; \hspace{1mm} (u_1,...,u_{D-1}) \mapsto S^x(u_1,...,u_{D-1})$ with semi-analytic but at least once differentiable functions $S^x(u_1,...,u_{D-1})$ and $S^x(0,...,0) = x$, and define $S^x_{\epsilon}(u_1,...,u_{D-1}) := S^x(\epsilon u_1,...,\epsilon u_{D-1})$. We find that (\ref{eq:fluxes}) becomes (with the choice $n_{IJ} = \delta^K_{[I}\delta^L_{J]}$)
\be
	\frac{1}{\epsilon^{(D-1)}}\pi^{IJ}(S^x_{\epsilon}) &=& \frac{1}{\epsilon^{(D-1)}} \int_{(-\epsilon/2,\epsilon/2)^{D-1}} du_1...du_{D-1} \epsilon_{aa_1...a_{D-1}} (\partial S^{xa_1}/\partial u_1)(u_1,...,u_{D-1})\times ... \nonumber \\ &\m& \times (\partial S^{xa_{D-1}}/\partial u_{D-1})(u_1,...,u_{D-1})\hspace{2mm} \pi^{aIJ}(S^x(u_1,...,u_{D-1})) \nonumber \\
	&=& n_a(S) \pi^{aIJ}(x) + O(\epsilon) 
\ee 
with $n_a(S) = \epsilon_{aa_1...a_{D-1}} (\partial S^{xa_1}/\partial u_1)(0,...,0)\times ... \times (\partial S^{xa_{D-1}}/\partial u_{D-1})(0,...,0)$, from which the claim follows. Now, similar to the treatment of the area operator in section \ref{sec:Area}, we just plug in the known quantisation of the electric fluxes and hope to get a well-defined constraint operator in the end. Using the regularised action of the flux vector fields on cylindrical functions (\ref{eq:FluxAction}), we find for a representative $f_{\gamma_{SS'}}$ of $f \in \text{Cyl}^2(\overline{\mathcal{A}})$ on a graph $\gamma_{SS'}$ adapted to both $S^x$ and $S'^x$,
\be
	\hat{C}_{\overline{M}}(S^x,S'^x)_{\gamma_{SS'}}\left[f_{\gamma_{SS'}}\right] &:=& \lim_{\epsilon,\epsilon' \rightarrow 0} \frac{1}{\epsilon^{(D-1)}\epsilon^{'(D-1)}} \epsilon_{IJKL\overline{M}} \hat{Y}^{IJ}_{\gamma_{SS'}}(S^x_{\epsilon}) \hat{Y}^{KL}_{\gamma_{SS'}}(S'^x_{\epsilon'}) [f_{\gamma_{SS'}}]  \nonumber \\
	&=& \lim_{\epsilon,\epsilon' \rightarrow 0} \frac{1}{\epsilon^{(D-1)}\epsilon^{'(D-1)}} \epsilon_{IJKL\overline{M}} \hspace{1mm}
	  \sum_{e \in E(\gamma_{SS'}); b(e) = x} \sum_{e' \in E(\gamma_{SS'}); b(e') = x} \nonumber \\ &\m& \epsilon(e,S^x) \epsilon(e',S'^x) R^{IJ}_e R^{KL}_{e'} f_{\gamma_{SS'}} \nonumber \\
	&=:& \lim_{\epsilon,\epsilon' \rightarrow 0} \frac{1}{\epsilon^{(D-1)}\epsilon^{'(D-1)}} \hat{\tilde{C}}_{\overline{M}}(S^x, S'^x)_{\gamma_{SS'}} [f_{\gamma_{SS'}}] \text{.} 
\ee
The flux vector fields only act locally on the intersection points $e\cap S$, $e \in E(\gamma_{SS'})$. Therefore, in the second line we used that for small surfaces $S^x_{\epsilon}$, $S'^x_{\epsilon'}$, the action of the constraint will be trivial expect for $x$ (and of course only non-trivial if $x$ is in the range of $\gamma_{SS'}$), thus independent of $\epsilon$. In the limit $\epsilon, \epsilon' \rightarrow 0$ the expression in the last line of the above calculation clearly diverges except for $\hat{\tilde{C}} f = 0$, where the whole expression vanishes identically. Since the kernels of the constraint operators $\hat{C}$ and $\hat{\tilde{C}}$ coincide, we can work with the latter and propose the constraint (omitting the $\sim$ again)

\be \hat{C}^{\overline{M}}(S,S',x)_{\gamma} p^*_{\gamma} f_{\gamma} 	&=& p^*_{\gamma_{SS'}} \epsilon^{IJKL\overline{M}} \sum_{e,e' \in \{e'' \in E(\gamma_{SS'}), b(e'') = x \} } \epsilon(e,S^v) \epsilon(e',S'^v) R_{IJ}^e R_{KL}^{e'} p^*_{\gamma_{SS'}\gamma} f_{\gamma} \nonumber \\
							&=& p^*_{\gamma_{SS'}} \epsilon^{IJKL\overline{M}} \left( R_{IJ}^{up} - R_{IJ}^{down} \right) \left( R_{KL}^{up'} - R_{KL}^{down'} \right) p^*_{\gamma_{SS'}\gamma} f_{\gamma} \text{,} \label{eq:QuantSimp}
\ee
where $R_{IJ}^{up(')} := \sum_{e \in E(\gamma_{SS'}), b(e) = x, \epsilon(e,S(')) = 1} R_{IJ}^e $ and similar for $R_{IJ}^{down(')}$. In the following, will drop the superscript $\m^x$ for the surfaces for simplicity.

The proof that the family $\hat{C}_{\gamma}^{\overline{M}}(S,S',x)$ is consistent and defines a vector field $\hat{C}^{\overline{M}}(S,S',x)$ on $\overline{\mathcal{A}}$ follows from the consistency of $\hat{Y}_n(S)$. To see that the operator is essentially self-adjoint, let $\mathcal{H}^0_{\gamma,\vec{\pi}}$ be the finite-dimensional Hilbert subspace of $\mathcal{H}^0$ given by the closed linear span of spin network functions over $\gamma$ where all edges are labelled with the same irreducible representations given by $\vec \pi$, $\mathcal{H}^0 = \overline{\oplus_{\gamma,\vec\pi} \mathcal{H}^0_{\gamma, \vec \pi} }$. Given any surfaces $S$, $S'$ we can restrict the sum over graphs to adapted ones since we have $\mathcal{H}^0_{\gamma,\vec\pi} \subset \mathcal{H}^0_{\gamma_{SS'},\vec\pi'}$ for the choice $\pi'_{e'} = \pi_{e}$ with $E(\gamma_{SS'}) \ni e' \subset e \in E(\gamma)$. Since  $\hat{C}^{\overline{M}}(S,S',x)$ preserves each $\mathcal{H}^0_{\gamma,\vec\pi}$, its restriction is a symmetric operator on a finite-dimensional Hilbert space, therefore self-adjoint. To see that it is symmetric, note that the right hand side of the first line of (\ref{eq:QuantSimp}) consists of right-invariant vector fields which commute. This is obvious for the summands with vector fields acting on distinct edges $e \neq e'$, and for $e = e'$ note that $\left[R^e_{IJ}, R^e_{KL}\right]$ is antisymmetric in $(IJ) \leftrightarrow (KL)$ and thus vanishes if contracted with $\epsilon^{IJKL\overline{M}}$. Now it is straight forward to see that  $\hat{C}^{\overline{M}}(S,S',x)$  itself is essentially self-adjoint.

Note that we did not follow the standard route to quantise operators, which would be to adjust the density weight of the simplicity constraint to be $+1$ (in its current form it is $+2$) and quantise it using the methods in \cite{ThiemannQSD5}. 
Rather, the quantisation displayed above parallels the quantisation of the (square of the) 
area operator in 3+1 dimensions and indeed we could have considered 
$\int d^{D-1}u\;\sqrt{| n^S_a n^S_b S^{ab}_{\overline{M}}|}$ for arbitrary surfaces $S$ and 
would have arrived at the above expression in the limit that $S$ shrinks to a point without
having to take away the regulator $\epsilon$ (the dependence on two rather than one 
surface can be achieved, to some extent, by an appeal to the polarisation identity). If 
we would have quantised it using the standard route then   
it would be necessary to have access to the volume operator. We will see in section \ref{sec:VolOp} that for the derivation of the volume operator in certain dimensions in the form we propose, which is a generalisation of the $3+1$ dimensional treatment, we need the above simplicity constraint operator to cancel some unwanted terms. Of course, there might be other proposals for volume operators which can be defined in any dimension without using the simplicity constraint. Still, the quantisation of the simplicity constraint presented here will (1) give contact to the simplicity constraints used in spin foam models and (2) enable us to solve the constraint in any dimension when acting on edges. Its action on the vertices, i.e. the requirements on the intertwiners, is more subtle and we propose to treat it using the Master constraint method. We will first present the action on edges and afterwards derive a suitable Master constraint. For following calculations, note that we always can adapt a graph to a finite number of surfaces. Furthermore, it is understood that all surfaces intersect $\gamma'$ in one point only (we may always shrink the surfaces until this is true).

\subsubsection{Edge Constraints and their Solution}
The action of the quantum simplicity constraint at an interior point $x$ of an analytic edge $e = e_1 \circ (e_2)^{-1}$ for both surfaces $S$, $S'$ not containing $e$ (otherwise the action is trivial) is given by
\be \hat{C}^{\overline{M}}(S,S',x)  p^*_{\gamma} f_{\gamma} 	&=& \pm p^*_{\gamma_{SS'}} \epsilon^{IJKL\overline{M}} \left( R_{IJ}^{e_1} - R_{IJ}^{e_2} \right) \left( R_{KL}^{e_1} - R_{KL}^{e_2} \right) p^*_{\gamma_{SS'}\gamma} f_{\gamma} \nonumber \\
 							&=& \pm p^*_{\gamma_{SS'}} 2 \epsilon^{IJKL\overline{M}} \left( R_{IJ}^{e_1} - R_{IJ}^{e_2} \right) R_{KL}^{e_1} p^*_{\gamma_{SS'}\gamma} f_{\gamma}  \nonumber \\
 							&=& \pm p^*_{\gamma_{SS'}} 2 \epsilon^{IJKL\overline{M}} R_{KL}^{e_1} \left( R_{IJ}^{e_1} - R_{IJ}^{e_2} \right) p^*_{\gamma_{SS'}\gamma} f_{\gamma}  \nonumber \\
						 	&=& \pm p^*_{\gamma_{SS'}} 4 \epsilon^{IJKL\overline{M}} R_{IJ}^{e_1} R_{KL}^{e_1} p^*_{\gamma_{SS'}\gamma} f_{\gamma} \text{,}
\ee
where the sign is $+$ if the orientation of the two surface $S$, $S'$ with respect to $e$ coincides and $-$ otherwise. In the second and fourth step we used gauge invariance at the vertex $v$ of an adapted graph, $\left[\sum_{e \in E(\gamma);\hspace{1mm} v=b(e)} R_{IJ}^e \right] f_{\gamma_{SS'}} = 0$, and in the third step we used that $\left[ R^{e_1}, R^{e_2} \right]=0$. This leads to the requirement on the generators of SO$(D+1)$ for all edges
\be \tau_{[IJ} \tau_{KL]} = 0 \text{.} \ee
The so-called simple representations of SO$(D+1)$ satisfying this constraint were classified in \cite{FreidelBFDescriptionOf}. Irreducible simple representations are given by homogeneous harmonic polynomials $\mathcal{H}_N^{(D+1)}$ of degree $N$, in any dimension labelled by one positive integer $N$. In this sense, there is a similarity between the simple representations of SO$(D+1)$ and the representations of SO$(3)$ (which all can be thought of as being simple). In particular, for $D+1 = 4$ we obtain the well-known simple representations of SO$(4)$ used in spin foams labelled by $j^+ = j = j^-$.

The commutator with gauge transformations at an interior point $x$ of an analytic edge $e = e_1 \circ (e_2)^{-1}$ ($e_1$, $e_2$ outgoing at $x$) yields, analogously to the classical calculation,
\be &\mbox{}& \left[ \hat{G}_{\gamma_{SS'}}[\Lambda], \hat{C}^{\overline{M}}(S,S',x)_{\gamma_{SS'}} \right] \nonumber \\
 &=& \pm \Lambda^{AB}(x) \epsilon^{IJKL\overline{M}} \left[ \left(R^{e_1}_{AB} + R^{e_2}_{AB}\right),  \left(R^{e_1}_{IJ} - R^{e_2}_{IJ}\right)  \left(R^{e_1}_{KL} - R^{e_2}_{KL}\right) \right] \nonumber \\
 &=& \pm \left\{\Lambda^{AB}(x) \epsilon^{IJKL\overline{M}} \left[R^{e_1}_{AB}, R_{IJ}^{e_1} R_{KL}^{e_1} - 2 R_{IJ}^{e_1} R_{KL}^{e_2} \right] + \left( e_1 \leftrightarrow e_2 \right)\right\} \nonumber \\ 
 &=& \pm \sum_{i=1}^{D-3} \Lambda^{M_i}\m_{M'_i}(x) \epsilon^{IJKLM_1...M_{i-1}M'_iM_{i+1}...M_{D-3}} \hspace{1mm} \left(R_{IJ}^{e_1} R_{KL}^{e_1} - 2 R_{IJ}^{e_1} R_{KL}^{e_2} + R_{IJ}^{e_2} R_{KL}^{e_2} \right) \nonumber \\ 
	 &=& \sum_{i=1}^{D-3} \Lambda^{M_i}\m_{M'_i}(x) \hspace{1mm} \hat{C}^{M_1...M_{i-1}M'_iM_{i+1}...M_{D-3}}(S,S',x) \text{.} \label{eq:simplicityrotated}\ee
Two constraints acting at the same interior point $x$ of an edge $e = e_1 \circ (e_2)^{-1}$ commute weakly. Using the gauge invariance of $Cf$ if $f$ is gauge invariant, we find
\be &\m& \left[\hat{C}^{\overline{M}}(S,S',x), \hat{C}^{\overline{N}}(S'',S''',x')\right] p^{*}_{\gamma} f_{\gamma} \nonumber \\
	&\approx& \pm 16 p^{*}_{\gamma} \delta_{x,x'} \epsilon^{IJKL\overline{M}} \epsilon^{OPQR\overline{N}} \left[ R^{e_1}_{IJ}R^{e_1}_{KL}, R^{e_1}_{OP} R^{e_1}_{QR} \right] f_{\gamma} + \mathcal{O}(\hat{C}f_{\gamma}) +\mathcal{O}(\hat{G}f_{\gamma}) \nonumber \\
	&\sim& p^{*}_{\gamma} \delta_{x,x'} \left(\epsilon R^{e_1} \cdot \hat{C}^{e_1,rot} + \hat{C}^{e_1,rot}  \cdot \epsilon R^{e_1}\right)f_{\gamma} \nonumber \\  &\sim& p^{*}_{\gamma} \delta_{x,x'} \left(\epsilon R^{e_1} \cdot \hat{C}^{e_1,rot} + [\hat{C}^{e_1,rot}, \epsilon R^{e_1}] + \epsilon R^{e_1} \cdot \hat{C}^{e_1,rot}\right)f_{\gamma} \nonumber \\ 
	&\sim& p^{*}_{\gamma} \delta_{x,x'} \left(2 \epsilon R^{e_1} \cdot  \hat{C}^{e_1,rot} + \epsilon \cdot \hat{C}^{e_1,rot,rot}\right)f_{\gamma} \approx 0 \text{,} \ee
which can be seen by the fact that the simplicity on an edge is quadratic in the rotation generator $R^{e_1}$ on that edge, and we used the notation 
\be \sum_{i=1}^{D-3} \Lambda^{M_i}\m_{M'_i} \hspace{1mm} \epsilon^{ABCDM_1...M_{i-1}M'_iM_{i+1}...M_{D-3}} R^e_{AB}R^e_{CD} =: \Lambda \cdot \hat{C}^{e,rot} \ee
for a simplicity with a infinitesimal rotation acting on the multi-index $\overline{M}$ (cf. (\ref{eq:simplicityrotated})). Here, we chose a graph $\gamma$ adapted to all four surfaces $S$, $S'$, $S''$, $S'''$. Note that classically, the Poisson bracket of two simplicity constraints vanishes strongly, whereas in the quantum theory this is only true in a weak sense. Still, the simplicity constraints acting on an edge are thus non-anomalous and can be solved by labelling all edges by simple representations of SO$(D+1)$.

\subsubsection{Vertex Master Constraint}
When acting on a node then, like the off-diagonal constraints in spin foam models, the simplicity constraints will not (weakly) commute anymore. Therefore, we are not allowed to introduce these constraints strongly and have the options of either trying to implement them weakly \cite{EngleTheLoopQuantum} or using a Master constraint. We will follow the latter route and give a proposal of how to construct a Master constraint of the simplicity constraints at the nodes. To reduce complexity, we try to find a both necessary and sufficient set of simple ``building blocks'' of the simplicity constraint at the node and construct a Master constraint using these. Considering (\ref{eq:QuantSimp}), an obviously sufficient set of building blocks at the vertex $v$ is given by
\be R^e_{[IJ}R^{e'}_{KL]}f_{\gamma}=0 \hspace{5mm} \forall e,e' \in \{e'' \in E(\gamma); v = b(e'')\}  \text{.} \ee
For necessity, we have to prove that we can choose surfaces in such a way that these building blocks follow. Note that it has already been shown in \cite{SahlmannIrreducibilityOfThe} that all right invariant vector fields $R^e$ for single edges $e$ can be generated by the $Y(S)$, but the construction involves commutators of the fluxes. Since the simplicity constraints acting on vertices are anomalous, we cannot use commutators in our argument. Instead, we will construct the right invariant vector fields $R^e$ by using linear combinations of fluxes only. To this end, we will prove the following lemma: 

\begin{lemma} \mbox{}\\
For each edge $e \in E(v)$ at the vertex $v$ we can always choose two surfaces $S$, $\tilde{S}$, such that the orientations with respect to $S$, $\tilde{S}$ of all edges but $e$ coincide. 
\end{lemma}
The intuitive idea of how to find these surfaces is to start with a surface containing the edge $e$ while intersecting all other edges $e' \in E(v), e' \neq e$ transversally, and then slightly distort this surface in the two directions ``above'' and ``below'' defined by the surface, such that the edge $e$ in consideration is once above and once below the surface, while the orientations of all other edges with respect to the surfaces remain unchanged, in particular none of them lies inside the surfaces. When subtracting the flux vector fields corresponding to the two distorted surfaces, all terms will cancel except the terms involving $R^e$.

\begin{proof}
To prove the statement above, two cases have to be distinguished: (a) the case where no $e' \in E(v)$ is (a segment of) the analytic extension through $v$ of the edge $e$ and (b) the case where $e$ has a partner $\tilde{e}$ which is a analytic extension of $e$ through $v$.

\textit{Case (a):}
The construction of the surface $S_{v,e}$ with the following properties 
\begin{enumerate}
\item $s_e \subset S_{v,e}$ for some beginning segment $s_e$ of $e$, and the other edges $e' \in E(v), e' \neq e$ intersect $S_{v,e}$ transversally in $v$.
\item For $e' \in E(v), e' \neq e$: $e'\cap S_{v,e} = v$, and for $e' \notin E(v)$, $e'\cap S_{v,e} = \emptyset$. 
\end{enumerate}
is given in \cite{SahlmannIrreducibilityOfThe} and we summarise the result shortly. An analytic surface (edge) is completely determined by its germ $[S]_v$ ($[e]_v$)
\be S(u_1,...,u_{D-1}) &=& \sum_{m_1,...,m_{D-1}=0}^{\infty} \frac{u_1^{m_1} ... u_{D-1}^{m_{D-1}}}{m_1! ... m_{D-1}!} S^{(m_1,...,m_{D-1})}\left(0,...,0\right) \text{,} \nonumber \\
e(t) &=& \sum_{n=0}^{\infty} \frac{t^n}{n!}e^{(n)}(0) \text{.} 
\ee
To ensure that $s_e \subset S_{v,e}$, we just need to choose a parametrisation of $S$ such that $S(t,0,...,0) = e(t)$ which fixes the Taylor coefficients $S^{(m,0,...,0)}(0,...,0) = e^{(m)}(0)$. For the finite number $k = |E(v)| - 1$ of remaining edges at $v$, we can now use the freedom in choosing the other Taylor coefficients to assure that there are no (beginning segments of) other edges contained in $S_{v,e}$ \cite{SahlmannIrreducibilityOfThe}. In particular, only a finite number of Taylor coefficients is involved.  

Now we state that the intersection properties of a finite number of transversal edges at $v$ with any (sufficiently small) surface $S$ are already fixed by a finite number of Taylor coefficients of $S$. We will discuss the case $D=3$ for simplicity, higher dimensions are treated analogously. Locally around $v$ we may always choose coordinates such that the surface is given by $z = 0$, $S(x,y) = (x,y,0)$. The edge $e$ contained in the surface is given by $e(t) = (x(t),y(t),0)$ and for any transversal edge at $v$ we find $e'(t) = (x'(t), y'(t), z'(t))$ where $z'(t) = \frac{t^{n-1}}{(n-1)!} z^{'(n-1)}(0) + \mathcal{O}(t^{n})$, and $n < \infty$ since otherwise $e'$ would be contained in $S$. The sign of the lowest non-vanishing Taylor coefficient $z^{'(n-1)}(0)$ determines if the edge is ``up''- or ``down''-type locally. Set $N = \max_{e' \in E(v), e' \neq e}{(n)}$, and obviously $N < \infty$. Thus, we can e.g. by modifying $S^{(N,0)}(0,0)$ choose the surface $\tilde{S}(x,y) = (x,y, \pm x^N)$, which locally has the same intersection properties with the edges $e' \in E(v), e' \neq e$ and certainly does not contain $e$ anymore.

Coming back to the general case considered before, there always exists $N < \infty$ such that we can change $S^{(N,0,...,0)}(0,...,0)$ without modifying the intersection properties of any of the edges $e' \in E(v), e' \neq e$, in particular the ``up''- or ``down''-type properties are unaffected. However, the edge $e$ no longer is of the inside type, but becomes either ``up'' or ``down'' (depending on whether $S^{(N,0,...,0)}(0,...,0)$ is scaled up or down and on the orientation of $S$). In general, new intersection points $v' \in E(v) \cap S, v' \neq v$ may occur when modifying the surface in the above described way, but we may always make $S$ smaller to avoid them.

Now choose a pair of surfaces $S$, $\tilde{S}$ for the edge $e$ such that it is once ``up''- and once ``down''-type to obtain the desired result
\be \left[\hat{Y}_{IJ}(S) - \hat{Y}_{IJ}(\tilde{S})\right] p^{*}_{\gamma} f_{\gamma} = 2 p^{*}_{\gamma} R^e_{IJ} f_{\gamma} \text{.} \label{eq:Lemma} \ee

\textit{Case (b):}
In the case that there is a partner $\tilde{e}$ which is a analytic continuation of $e$ through $v$, we cannot construct an analytic surface (without boundary) $S_{v,e}$ containing a beginning segment of $e$ and not containing a segment of $\tilde{e}$. However, we can construct an analytic surface $S_{v,\left\{e,\tilde{e}\right\}}$ containing (beginning segments of) $e$, $\tilde{e}$ and sharing the remaining properties with $S_{v,e}$ above. The method is the same as in case (a) \cite{SahlmannIrreducibilityOfThe}. Again, there always exists $N < \infty$ such that we can change $S^{(N,0,...,0)}(0,...,0)$ without modifying the intersection properties of any of the edges $e' \in E(v), e' \neq \{e,\tilde{e}\}$, and such that both edges $e$, $\tilde{e}$ become either ``up'' or ``down''-type. Moreover, if we choose $N$ even, then $e$, $\tilde{e}$ will be of the same type with respect to the modified surface, while for $N$ odd one edge will be ``up'' and its partner will be ``down''. Calling the modified surface $S$ for $N$ even and $\tilde{S}$ for $N$ odd, we find with the same calculation (\ref{eq:Lemma}) as in case (a) the desired result. \\
This furnishes the proof of the above lemma\footnote{This also establishes that the right invariant vector fields $R^e_{IJ}$ are not only contained in the Lie algebra generated by the flux vector fields $\hat{Y}(S)$, but are already contained in the flux vector space, which to the best of our knowledge has not been shown.}.
\end{proof}
Choosing the surfaces as described above, we find that the following linear combination
\begin{eqnarray} 
&&\frac{1}{4} \left(\hat{C}^{\overline{M}}(S,S',x) - \hat{C}^{\overline{M}}(\tilde{S},S',x) - \hat{C}^{\overline{M}}(S,\tilde{S}',x) + \hat{C}^{\overline{M}}(\tilde{S},\tilde{S}',x)\right) p^*_{\gamma}f_{\gamma} \nonumber \\ &=& p^*_{\gamma} \epsilon^{IJKL{\overline{M}}} R^e_{IJ}R^{e'}_{KL} f_{\gamma} 
\end{eqnarray}
proves the necessity of the building blocks. Using the fact that the edge representations are already simple, we can rewrite the building blocks as
\be  R^e_{[IJ} R^{e'}_{KL]} f_{\gamma} &=& \frac{1}{2} \left[(R^e_{[IJ} + R^{e'}_{[IJ})(R^e_{KL]} + R^{e'}_{KL]}) - R^e_{[IJ} R^{e}_{KL]} - R^{e'}_{[IJ} R^{e'}_{KL]}\right] f_{\gamma} \nonumber \\ &=& \frac{1}{2} (R^e_{[IJ} + R^{e'}_{[IJ})(R^e_{KL]} + R^{e'}_{KL]}) f_{\gamma} =: \frac{1}{2} \Delta^{ee'}_{IJKL} f_{\gamma} \text{.}  \label{eq:allsimplicities} \ee

We proceed by showing that the building blocks are anomalous, starting with the case $D=3$. We calculate for $e \neq e' \neq e'' \neq e$
\be
\left[ \epsilon^{IJKL} \Delta^{ee'}_{IJKL},  \epsilon^{ABCD} \Delta^{e'e''}_{ABCD} \right] \sim \delta_{IJK}^{ABC} (R_{e''})_{AB} (R_{e})^{IJ} (R_{e'})^K\m_C\text{,} \label{eq:anomaly}
\ee
where we used the notation $\delta^{I_1...I_n}_{J_1...J_n} := n! \; \delta^{I_1}_{[J_1} \delta^{I_2}_{J_2} ... \delta^{I_n}_{J_n]}$. To show that this expression can not be rewritten as  a linear combination of the of building blocks (\ref{eq:allsimplicities}), we antisymmetrise the indices $[ABIJ]$, $[ABKC]$ and $[IJKC]$ and find in each case that the result is zero. Therefore, a simplicity building block can not be contained in any linear combination of terms of the type (\ref{eq:anomaly}). 
For $D>3$, we have
\be
\left[ \epsilon^{IJKL\overline{M}} \Delta^{ee'}_{IJKL},  \epsilon^{ABCD\overline{E}} \Delta^{e'e''}_{ABCD} \right] \sim \delta_{IJK\overline{M}}^{ABC\overline{E}} (R_{e''})_{AB} (R_{e})^{IJ} (R_{e'})^K\m_C \label{eq:VertexSimplicityAnomaly} \text{.}
\ee
Choosing $\overline{M} = \overline{E}$ fixed, the anomaly is the same as above. A short remark concerning the terminology ``anomaly'' is in order at this place. Normally, the term anomaly denotes that a certain classical structure, e.g. the constraint algebra, is not preserved at the quantum level, e.g. by factor ordering ambiguities. The non-commutativity of the simplicity constraints however can already be seen at the classical level when using holonomies and fluxes as basic variables. Thus, one could argue that it would be more precise to talk of a quantisation of second class constraints. On the other hand, since the holonomy-flux algebra is an integral part of the quantum theory and at the classical level it would be perfectly fine to use a non-singular smearing, we will nevertheless use the term anomaly to describe this phenomenon.

Independently of the terminology chosen, we cannot quantise the simplicity constraints acting on vertices using the Dirac procedure since this will lead to the additional constraints  (\ref{eq:VertexSimplicityAnomaly}) being imposed. The unique solution to these constraints has been worked out in \cite{FreidelBFDescriptionOf} and is given by the Barrett-Crane intertwiner in four dimensions and a higher-dimensional analogue thereof. Several options are at our disposal at this point. Looking back at our companion paper \cite{BTTII}, one could try to gauge unfix this second class system to obtain a first class system subject to only a subset of the vertex simplicity constraints. In this process, one would have to pick out a first class subset of the simplicity constraints which has a closing algebra with the remaining constraints. The construction of a possible choice of such a subset  is discussed in our companion paper \cite{BTTV}. While the proposed subset is first class with respect to the other constraints, it suffers from the fact that the choice is based on a certain recoupling scheme and that a different choice of the recoupling scheme results in a different first class subset. This is not a problem for the theory itself, but it seems problematic when constructing a unitary map to SU$(2)$ spin networks in four dimensions, as discussed in \cite{BTTV}. Another possibility is the construction of a Dirac bracket, which however would result in a non-commuting connection and the non-applicability of the LQG quantisation methods. The use of a weak implementation in the sense of Gupta and Bleuler is discussed in our companion paper \cite{BTTV}. While the results obtained in the context of the EPRL spin foam model can be also used in the canonical theory (up to certain subtleties discussed in \cite{BTTV}), they rely on specific properties of SO$(4)$ which do not extend to higher dimensions.

While equivalent at the classical level, the master constraint introduced in \cite{ThiemannThePhoenixProject} allows to quantise also second class constraints by a strong operator equation. Due to the second class nature, one expects the master constraint operator to have an empty kernel or at least a kernel which is too small to describe the physical Hilbert space. Since we know that the Barrett-Crane intertwiner is a solution to the strong imposition of all vertex simplicity constraints, we are in the second case. In order to find a larger kernel of the master constraint, one modifies it by adding terms to it which vanish in the classical limit, i.e. performs $\hbar$-corrections. The merits of this procedure are exemplified by the construction of the EPRL intertwiner \cite{EngleLoopQuantumGravity} in four dimensions, which results from a master constraint for the linear simplicity constraint upon $\hbar$-corrections. A simplification arising in the treatment of the linear simplicity constraints, see e.g. \cite{EngleLoopQuantumGravity} our companion paper \cite{BTTV}, is that they act individually on every edge connected to the intertwiner. On the other hand, the quadratic constraints act on pairs of edges and the resulting algebraic structure of the master constraints is thus very different. Since we are not aware of a suitable solution for the quadratic vertex master simplicity constraint, we will contend ourselves by giving a definition of this constraint operator. The task remaining for solving the vertex simplicity master constraint operator is thus to find a proper $\hbar$-correction which results in a physical Hilbert space with the desired properties, e.g. that there exists a unitary map to SU$(2)$ spin networks in four dimensions.

A general simplicity Master constraint is given by
\be
\boldsymbol{\hat{M}}_v p^{*}_{\gamma} f_{\gamma} = p^{*}_{\gamma} \sum_{e,e',e'',e''' \in E(v)} c_{ee'}^{e''e'''}\m_{IJKL}^{MNOP} \Delta^{ee'}_{IJKL} \Delta^{e''e'''}_{MNOP} f_{\gamma}
\ee
with a positive matrix $c_{ee'}^{e''e'''}\m_{IJKL}^{MNOP}$, which we will choose diagonal for simplicity, $c_{ee'}^{e''e'''}\m_{IJKL}^{MNOP} = \frac{1}{4!} c_{ee'} \delta_{(e}^{e''} \delta_{e')}^{e'''} \delta_{IJKL}^{MNOP}$. The diagonal elements $c_{ee'}$ can be chosen symmetric because of the symmetry of the building blocks. We choose $c_{ee'} = 1 \hspace{1mm} \forall \hspace{1mm} e, e', e\neq e'$ and $c_{ee} = 0$ since the edge representations are already simple, leading to the final version of the Master constraint we propose,
\be
\boldsymbol{\hat{M}}_v p^{*}_{\gamma} f_{\gamma}  = p^{*}_{\gamma} \sum_{e,e' \in E(v), e\neq e'} \Delta^{ee'}_{IJKL} \Delta^{ee'}_{IJKL}  f_{\gamma} \text{.}
\ee
Cylindrical consistency and essential self-adjointness follows analogously to the case of $C(S,S',x)$ in section \ref{sec:Simplicity1}. 

For the case of SO$(4)$, we can use the decomposition in self-dual and anti-selfdual generators to find that $\epsilon^{IJKL} R^e_{IJ}R^{e'}_{KL} = \vec{J}^e_+ \cdot \vec{J}^{e'}_+ - \vec{J}^e_-\cdot \vec{J}^{e'}_-$, which implies 
\be \epsilon^{IJKL} \Delta^{ee'}_{IJKL} = \left(\vec{J}^e_+ + \vec{J}^{e'}_+\right)\cdot\left(\vec{J}^e_+ + \vec{J}^{e'}_+\right) - \left(\vec{J}^e_- + \vec{J}^{e'}_-\right)\cdot\left(\vec{J}^e_- + \vec{J}^{e'}_-\right) =: \Delta^{ee'}_+ - \Delta^{ee'}_-  \text{.} \ee
This leads to the Master constraint
\be
\boldsymbol{\hat{M}}_v p^{*}_{\gamma} f_{\gamma} = p^{*}_{\gamma} \sum_{e,e' \in E(v), e\neq e'} \left( \Delta^{ee'}_+\Delta^{ee'}_+ - 2\Delta^{ee'}_+\Delta^{ee'}_- + \Delta^{ee'}_-\Delta^{ee'}_- \right)  f_{\gamma}  \text{,}
\ee
where $+$ and $-$ now label independent copies of SO$(3)$. Thus, we can calculate the matrix elements of this constraint in a recoupling basis analogously to the standard LQG volume operator matrix elements \cite{ThiemannClosedFormulaFor}.

As mentioned before, alternative routes to deal with the vertex simplicity constraints will be the subject of \cite{BTTV}.

\subsection{Diffeomorphism Constraint}
The diffeomorphism constraint can again be treated in exact agreement with the $(3+1)$-dimensional case. To solve the diffeomorphism constraint, one proceeds as follows. Consider the set of smooth cylindrical functions $\Phi := \text{Cyl}^{\infty}(\overline{\mathcal{A}/\mathcal{G}})$ which can be shown to be dense in $\mathcal{H}^0$. By a distribution $\psi \in \Phi'$ on $\Phi$ we 
simply mean a linear functional on $\Phi$. 
The group average of a spin-network state $T_{\gamma,\vec{\Lambda},\vec{c}}$ is defined by the following well-defined distribution on $\Phi$
\be T_{[\gamma],\vec{\Lambda},\vec{c}} := \sum_{\gamma' \in [\gamma]} 
<T_{\gamma',\vec{\Lambda},\vec{c}},.> \text{,}\ee
where $[\gamma]$ denotes the orbit of $\gamma$ under smooth diffeomorphisms of $\sigma$ which preserve the analyticity of $\gamma$ including an average over the 
graph symmetry group (see, e.g.,  \cite{ThiemannQSD3} for technical details). Since we already solved the simplicity constraint on single edges, we can restrict attention to spin network states with edges labelled by simple SO$(D+1)$ representations, $\Lambda_e = (N_e,0,...)$. The group average $[f]$ of a general cylindrical function $f$ is defined by demanding linearity of the averaging procedure, i.e. first decompose $f$ into spin-network states and then average each of the spin-network states separately. An inner product for the diffeomorphism invariant Hilbert space can be constructed. We will not give details and refer the reader to \cite{AshtekarQuantizationOfDiffeomorphism, ThiemannQSD3}.
 
\section{Geometrical Operators}

\subsection{The $D-1$ Area Operator}
\label{sec:Area}
The area operator was first considered in \cite{SmolinRecentDevelopmentsIn} and defined 
mathematically rigorously  in the LQG representation in \cite{AshtekarQuantumTheoryOf1}. In \cite{ThiemannModernCanonicalQuantum}, the results of \cite{AshtekarQuantumTheoryOf1} are generalised for arbitrary dimension $D$. Using the classical identity $\pi^{aIJ} \pi^{b}\m_{IJ} = 2 q q^{ab}$, we can basically copy the treatment found there. Let S be a surface and $X: U_0 \rightarrow S$ the associated embedding, where $U_0$ is an open submanifold of $\mathbb{R}^{D-1}$. Then the area functional is given by
\be \text{Ar}[S] := \int_{U_0} d^{D-1}u \sqrt{det\left(\left[X^*q\right]\left(u\right)\right)} \text{.}\ee
Introduce $U_0 = \cup_{U \in \mathcal{U}} U$, a partition of $U_0$ by closed sets $U$ with open interior, $\mathcal{U}$ being the collection of these sets. Then the area functional can be written as the limit as $|U|\rightarrow \infty$ of the Riemann sum
\be \text{Ar}[S] := \sum_{U \in \mathcal{U}} \sqrt{\frac{1}{2} \pi_{IJ}(S_U)\pi^{IJ}(S_U)} \text{,}\ee
where $S_U = X(U)$ and $\pi_{IJ}(S_U)$ is the electric flux with choice $n^{IJ} = \delta^I_{[K}\delta^J_{L]}$, which has been quantised already. Let $f \in \text{Cyl}^2(\overline{\mathcal{A}})$, choose a representative $f_{\gamma}$ and, using the known action of the quantised electric fluxes, obtain as in the $(3+1)$-dimensional case
\be \widehat{\text{Ar}_{\gamma}}[S]p^*_{\gamma} f_{\gamma} = \kappa \hbar \beta p_{\gamma_S}^* \sum_{x \in \{e \cap S;  e \in E(\gamma_S)\}} \sqrt{- \frac{1}{2}\left\{ \sum_{e \in E(\gamma_S), x\in \partial e} \epsilon(e,S) R_{IJ}^e \right\}^2} p^*_{\gamma_S \gamma} f_{\gamma} \text{,}\ee
where $\gamma_S \succ \gamma$ is an adapted graph. The family of operators $\widehat{\text{Ar}_{\gamma}[S]}$ has dense domain $\text{Cyl}^2(\overline{\mathcal{A}})$. Its independence of the adapted graph follows from that of the electric fluxes. Moreover, the properties of the area operator like cylindrical consistency, essential self-adjointness and discreteness of the spectrum can be shown analogously to \cite{ThiemannModernCanonicalQuantum}. 

The complete spectrum can be derived using the standard methods. We use
\be \left\{ \sum_{e \in E(\gamma_S), x\in \partial e} \epsilon(e,S) R_{IJ}^e \right\}^2 &=& 2 \left( R_{IJ}^{x,up} \right)^2 + 2 \left( R_{IJ}^{x,down} \right)^2 - \left( R_{IJ}^{x,up} + R_{IJ}^{x,down} \right)^2 \nonumber \\
&=:& - \Delta^{up} - \Delta^{down} + \frac{1}{2} \Delta^{up+down} \text{,}
\ee
where the $\Delta$s are mutually commuting primitive Casimir operators of SO$(D+1)$. Thus their spectrum is given by the Eigenvalues $\lambda_{\pi} > 0$. We have to distinguish the cases $D+1= 2n$ even, $\mathbb{N} \ni n \geq 2$ and $ D+1 = 2n+1$ odd, $n \in \mathbb{N}$. In a representation of SO$(D+1)$ with highest weight $\Lambda = (n_1,...,n_n)$, $n_i \in \mathbb{N}_0$, we find for the eigenvalues of the Casimir\footnote{Note that $R^{IJ} = 1/2 X^{IJ}$, such that $X^{IJ}$ fulfil the standard Lie algebra relations without the factor $1/2$ appearing in (\ref{eq:LieAlgebraRelations}).} $\Delta := -\frac{1}{2} X_{IJ} X^{IJ}$
\be \Delta v_{\Lambda} &:=& \lambda_{\pi_{\Lambda}} v_{\Lambda} = \left[\sum_{i=1}^n f_i^2 + 2\sum_{j=2}^n\sum_{i<j} f_i\right] v_{\Lambda} \hspace{5mm} \text{for SO$(2n)$,} \nonumber \\ 
		\Delta v_{\Lambda} &:=& \lambda_{\pi_{\Lambda}} v_{\Lambda} = \left[\sum_{i=1}^n f_i^2 + 2\sum_{j=2}^n\sum_{i<j} f_i + \sum_{i=1}^n f_i\right] v_{\Lambda} \hspace{5mm} \text{for SO$(2n+1)$,} \label{eq:Casimir}
\ee
where we used the following notation
\be f_i &=& \sum_{j=i}^{n-2}n_j + \frac{n_{n-1}+n_n}{2}, \hspace{1mm} i\leq (n-2); \hspace{3mm} f_{n-1} = \frac{n_{n-1}+n_n}{2}; \hspace{3mm} f_{n} = \frac{n_n - n_{n-1}}{2} \hspace{5mm} \text{for SO$(2n)$,}\nonumber \\
		f_i &=& \sum_{j=i}^{n-1}n_j + \frac{n_n}{2}, \hspace{1mm} i\leq (n-1); \hspace{3mm} f_{n} = \frac{n_n}{2} \hspace{5mm} \text{for SO$(2n+1)$,} 
 \ee
such that $f_1 \geq f_2 \geq ... \geq f_n$. Note that the above formulas hold for general irreducible Spin$(D+1)$ representations. Irreducible representations of SO$(D+1)$ are found by the restriction that all $f_i$ be integers. Denoting by $\Pi$ a collection of representatives of irreducible representations of SO$(D+1)$, one for each equivalence class, we find for the area spectrum
\be
\text{Spec}(\widehat{\text{Ar}}[S]) = \left\{\frac{\kappa \hbar \beta}{2} \sum_{n=1}^N \sqrt{2\lambda_{\pi_n^1} + 2\lambda_{\pi_n^2} - \lambda_{\pi_n^{12}}}; \hspace{1mm} N\in\mathbb{N}, \hspace{1mm} \pi^1_n,\pi^2_n,\pi^{12}_n \in \Pi, \hspace{1mm} \pi^{12}_n \in \pi^1_n \otimes \pi^2_n \right\} \text{.} \label{eq:AreaSpectrum}
\ee
Note that the above formulas (\ref{eq:Casimir}) significantly simplify if we restrict to simple representations, $\Lambda_0 = (N,0,0,...)$,
\begin{alignat}{3} \Delta v_{\Lambda_0} &= N(N+2n-2) v_{\Lambda_0} &=& N(N+D-1)v_{\Lambda_0} \hspace{5mm} &&\text{for SO$(2n)$,} \nonumber \\ 
		\Delta v_{\Lambda_0} &= N(N+2n+1-2) v_{\Lambda_0} &=& N(N+D-1)v_{\Lambda_0} \hspace{5mm} &&\text{for SO$(2n+1)$.} \label{eq:SimpleCasimir}
\end{alignat}
We cannot use this simplified expression for the SO$(D+1)$ Casimir operator in the general case (\ref{eq:AreaSpectrum}), since in the decomposition of a tensor product of irreducible simple representations usually non-simple representations will appear\footnote{For the tensor product of two irreducible simple representations of SO$(n)$ holds \cite{GirardiGeneralizedYoungTableaux, GirardiKroneckerProductsFor} (w.l.o.g. $M \geq N$) $\m{[M,0,..,0] \otimes [N,0,..,0] = \sum_{K=0}^N \sum_{L=0}^{N-K} [M+N-2K-L,L,0,..,0]}$.}, but we can use it for a single edge. When acting on a single edge $e = e_1 \circ (e_2)^{-1}$ intersecting $S$ transversally, we know that due to gauge invariance
\be \left\{ R_{IJ}^{e_1} - R_{IJ}^{e_2} \right\}^2 h_{e} = 4 \left(R^{e_1}_{IJ}\right)^2 h_{e} = - 2N(N+D-1) h_{e} \text{.}
\ee
The action of the area operator on a single edge $e$, $e\cap S \neq \emptyset$ is thus given by
\be
\widehat{\text{Ar}_e}[S] p^*_{e} h_{e} =  \kappa \hbar \beta \sqrt{N(N+D-1)} p^*_{e} h_{e} = 16 \pi \beta \left(l^{(D+1)}_p\right)^{D-1} \times \sqrt{N(N+D-1)} p^*_{e} h_{e} \text{,}
\ee
where $l^{(D+1)}_p := \sqrt[D-1]{\frac{\hbar G^{(D+1)}}{c^3}}$ is the unique length in $D+1$ dimensions, and $\kappa = 16 \pi G^{(D+1)}/c^3$ in any dimension, where $G^{(D+1)}$ denotes the gravitational constant. Note that for $D=3$, we find the factor $\sqrt{N(N+2)}$ in the area spectrum of an edge stemming from irreducible simple representations of SO$(4)$. Replace the non-negative integer $N$ labelling the weight by $N = 2j$, $j$ half integer, to find the factor $2 \sqrt{j(j+1)}$ of SO$(4)$ spin foam models, which coincides with the usual spacing in $(3+1)$-dimensional LQG,
\be \widehat{\text{Ar}_e}[S] p^*_{e} h_{e} =  2\kappa \hbar \beta \sqrt{j(j+1)} p^*_{e} h_{e} = 32 \pi \beta \left(l^{(D+1)}_p\right)^{D-1} \times \sqrt{j(j+1)} p^*_{e} h_{e} \text{.}\ee
In standard LQG, instead of the gauge group SO$(3)$ one extends to the double cover Spin$(3) \cong$ SU$(2)$ and allows also for half integer representations. Note that in our case, we cannot allow for general Spin$(D+1)$ representations at the edges, since the edge simplicity constraint is not satisfied in representations of Spin$(D+1)$ which are not as well representations of SO$(D+1)$, $D \geq 3$ \cite{FreidelBFDescriptionOf}.

\subsection{The Volume Operator}
\label{sec:VolOp}
The derivation of the volume operator is analogous to the treatment in \cite{ThiemannModernCanonicalQuantum} and requires only a slight adjustment. 

The volume of a region $R$ is classically measured by
\be  V(R):= \int_R d^Dx \, \sqrt{q} \text{,}  \ee 
where $\sqrt{q}$ has to be expressed in terms of the canonical variables. The derivation is performed for $\beta = 1$, the general result is obtained by multiplying the resulting operator by $\beta^{D/(D-1)}$. 

\subsubsection{$D+1$ Even}

Let $n=(D-1)/2$. Let $\chi_\Delta(p,x)$ be the characteristic function in the coordinate $x$ of a hypercube with centre $p$ spanned by the $D$ vectors $\vec{\Delta}^i := \Delta^i \vec{n}^i$, $i = 1, \ldots, D$, where $\vec{n}^i$ is a normal vector in the frame under consideration and which has coordinate volume $\text{vol} = \Delta^1 \ldots \ \Delta^D \det(\vec{n}^1, \ldots, \vec{n}^D)$ (we assume the vectors to be right-oriented). 
In other words, 
\be  \chi_\Delta(p,x) = \prod^D_{i=1} \Theta \left(\frac{\Delta^i}{2}-\left|  < n^i, x-p > \right| \right)  \ee 
where $< \cdot, \cdot >$ is the standard Euclidean inner product and $\Theta(y) = 1$ for $y>0$ and zero otherwise. We will use lower indices $(\Delta_I^1, \ldots, \Delta_I^D)$ to label different hypercubes. It will turn out to be convenient to label the $D$ edges appearing in the following formulae by $e, e_1, \ldots, e_n, e'_1, \ldots, e'_n$.

We consider the smeared quantity
\begin{eqnarray}
 & & \pi(p, \Delta_1, \ldots, \Delta_D)  \nonumber \\
  & = & \frac{1}{ \text{vol}(\Delta_1)  \ldots \text{vol}(\Delta_D) } \int_\sigma d^Dx_1 \ldots \int_\sigma d^Dx_D  \nonumber \\
  & &\chi_{\Delta_1}(p, x_1) \chi_{\Delta_2}(2p, x_1 +x_2) \ldots \chi_{\Delta_D}(Dp, x_1 + \ldots  + x_D) \nonumber \\
  & & \frac{1}{2D!} \epsilon_{a a_1 b_1 \ldots a_n b_n} \epsilon_{IJI_1 J_1 I_2 J_2 \ldots I_nJ_n} \pi^{aIJ} \pi^{a_1 I_1 K_1} \pi^{b_1 J_1} \m_{K_1} \ldots \pi^{a_n I_n K_n} \pi^{b_n J_n} \m_{K_n} \text{.}
\end{eqnarray}
Then it is easy to see that the classical identity
\be  V(R) = \lim_{\Delta_1 \rightarrow 0} \ldots \lim_{\Delta_D \rightarrow 0} \int_R d^Dp\,\left| \pi(p, \Delta_1 , \ldots, \Delta_D) \right|^{\frac{1}{D-1}}  \ee 
holds. The canonical brackets
\be  \left\{ A_{aIJ}(x), \pi^{bKL}(y) \right\} = 2 \delta^D(x-y) \delta_a^b \delta_I^{[K} \delta_J^{L]}  \ee 
give rise to the operator representation
\be  \hat{\pi}^{bKL} = -\frac{\hbar}{i} \frac{\delta}{\delta A_{bKL}}  \ee 
while the connection acts by multiplication.

Let a graph $\gamma$ be given. In order to simplify the notation, we subdivide each edge $e$ with endpoints $v, v'$ which are vertices of $\gamma$ into two segments $s, s'$ where $e = s \circ (s')^{-1}$ and $s$ has an orientation such that it is outgoing at $v'$. This introduces new vertices $s \cap s'$ which we will call pseudo-vertices because they are not points of non-semianalyticity of the graph. Let $E(\gamma)$ be the set of these segments of $\gamma$ but $V(\gamma)$ the set of true (as opposed to pseudo) vertices of $\gamma$. Let us now evaluate the action of 
\be 
\hat{\pi}^{aIJ}(p, \Delta) := \frac{1}{\text{vol}(\Delta)} \int_\Sigma d^Dx \, \chi(p,x) \hat{\pi}^{aIJ} 
\ee
on a function $f = p^*_\gamma f_\gamma$ cylindrical with respect to $\gamma$. We find ($e:[0,1] \rightarrow \sigma, t\rightarrow e(t)$ being a parametrisation of the edge $e$)
\be \hat{\pi}^{aIJ}(p, \Delta) f = \frac{i \hbar}{\text{vol}(\Delta)} \sum_{e \in E(\gamma)} \int_0^1 \chi_\Delta(p, e(t)) \dot{e}^a(t) \text{tr} \left( \left[ h_e(0,t) \tau^{IJ} h_e(t,1) \right]^T \frac{\partial}{\partial h_e(0,1)} \right) f_\gamma \text{.} \label{eq:pihat} 
\ee
Here we have used (1) the fact that a cylindrical function is already determined by its values on $\mathcal{A} / \mathcal{G}$ rather than $\overline{\mathcal{A} / \mathcal{G}}$ so that it makes sense to take the functional derivative, (2) the definition of the holonomy as the path-ordered exponential of $\int_e A$ with the smallest parameter value to the left, (3) $A= dx^a A_{aIJ} \tau^{IJ}$ where $\tau^{IJ} \in \text{so}(D+1)$ and we have defined (4) $\text{tr}(h^T \partial / \partial g) = h_{AB} \partial / \partial_{AB}$, $A,B,C,\ldots$ being SO$(D+1)$ indices. The state that appears on the right-hand side of (\ref{eq:pihat}) is actually well-defined, in the sense of functions of connections, only when $A$ is smooth for otherwise the integral over $t$ does not exist, see \cite{mourao1999} for details. However, as announced, we will be interested only in quantities constructed from operators of the form (\ref{eq:pihat}) and for which the limit of shrinking $\Delta \rightarrow 0$ to a point has a meaning in the sense of $\mathcal{H} = L_2(\overline{\mathcal{A} / \mathcal{G}}, d\mu_0)$ and therefore will not be concerned with the actual range of the operator (\ref{eq:pihat}) for the moment. 

We now wish to evaluate the whole operator $\hat{\pi}(p, \Delta^1, \ldots, \Delta^D)$ on $f$. It is clear that we obtain $D$ types of terms, the first type comes from all three functional derivatives acting on $f$ only, the second type comes from $D-1$ functional derivatives acting on $f$ and the remaining one acting on the trace appearing in (\ref{eq:pihat}), and so forth. 

The first term (type) is explicitly given by
\begin{alignat}{3}
   & \hat{\pi}(p, \Delta_1, \ldots, \Delta_D) f \\
   = & \frac{1}{2D!} \frac{(i \hbar)^D}{ \text{vol}(\Delta_1) \ldots \text{vol}(\Delta_D)} \epsilon_{a a_1 b_1 \ldots a_n b_n} \epsilon_{IJI_1 J_1 I_2 J_2 \ldots I_nJ_n} \int_{[0,1]^D} dt\,dt_1 \ldots dt_n \, dt'_1 \ldots dt'_n \sum_{e_1, \ldots, e_D \in E(\gamma)} \label{eq:volf} \nonumber \\
   & \chi_{\Delta_1}(p, x_1) \chi_{\Delta_2}(2p, x_1 +x_2) \ldots \chi_{\Delta_D}(Dp, x_1 + \ldots  + x_D) \dot{e}^a(t) \dot{e}_1^{a_1}(t_1) \ldots \dot{e}_n^{a_n}(t_n) \dot{e}'_1 \m^{b_1}(t'_1) \ldots \dot{e}'_n \m^{b_n}(t'_n) \nonumber \\
   & \text{tr} \left( \left[ h_e(0,t) \tau^{IJ} h_e(t,1) \right]^T \frac{\partial}{\partial h_{e}(0,1)} \right) \nonumber \\
   & \text{tr} \left( \left[ h_{e_1}(0,t_1) \tau^{I_1 K_1 } h_{e_1}(t_1,1) \right]^T \frac{\partial}{\partial h_{e_1}(0,1)} \right)  \text{tr} \left( \left[ h_{e'_1}(0,t'_1) \tau^{J_1} \m_{K_1} h_{e'_1}(t'_1,1) \right]^T \frac{\partial}{\partial h_{e'_1}(0,1)} \right) \ldots \nonumber\\
    & \text{tr} \left( \left[ h_{e_n}(0,t_n) \tau^{I_n K_n } h_{e_n}(t_n,1) \right]^T \frac{\partial}{\partial h_{e_n}(0,1)} \right)  \text{tr} \left( \left[ h_{e'_n}(0,t'_n) \tau^{J_n} \m_{K_n} h_{e'_n}(t'_n,1) \right]^T \frac{\partial}{\partial h_{e'_n}(0,1)} \right) f_\gamma \text{.} \nonumber
\end{alignat}
The other terms are vanishing due to either the same symmetry / anti-symmetry properties as in the usual treatment or the simplicity constraint in case the first derivative is involved. 

Given a $D$-tuple $e_1 \ldots e_D$ of (not necessarily distinct) edges of $\gamma$, consider the functions 
\be  x_{e_1, \ldots, e_D} (t_1, \ldots, t_D) := e_1(t_1) + \ldots + e_D(t_D) \text{.}  \ee 
This function has the interesting property that the Jacobian is given by 
\be  \det \left( \frac{\partial (x^1_{e_1, \ldots, e_D}, \ldots x^D_{e_1, \ldots, e_D})(t_1, \ldots, t_D)}{\partial (t_1, \ldots, t_D)} \right)  = \epsilon_{a_1 \ldots a_D} \dot{e}_1(t_1)^{a_1} \ldots \dot{e}_D(t_D)^{a_D}  \ee 
which is precisely the form of the factor which enters the integral (\ref{eq:volf}).

We now consider the limit $\Delta^1, \ldots, \Delta^D \rightarrow 0$. The idea is that all quantities in (\ref{eq:volf}) are meaningful in the sense of functions on smooth connections and thus limits of functions as $\Delta \rightarrow 0$ are to be understood with respect to any Sobolev topology. The miracle is that the final function is again cylindrical and thus the operator that results in the limit has an extension to all of $\overline{\mathcal{A} / \mathcal{G}}$. 

\begin{lemma} \mbox{}\\
For each $D$-tuple of edges $e_1, \ldots, e_D$ there exists a choice of vectors $\vec{n}^1_1, \ldots, \vec{n}^1_D, \vec{n}^2_1, \ldots, \vec{n}^D_D$ and a way to guide the limit $\Delta^1_1, \Delta^1_2, \ldots, \Delta^D_D \rightarrow 0$ such that
\be  \int_{[0,1]^D} \det \left( \frac{\partial x^a_{e_1, \ldots, e_D}}{\partial (t_1, \ldots, t_D)} \right) \chi_{\Delta_1}(p, e_1) \ldots \chi_{\Delta_D}(Dp, e_1 + \ldots e_D) \hat{O}_{e_1, \ldots, e_D}  \ee 
vanishes if

\renewcommand{\labelenumi}{(\alph{enumi})}

\begin{enumerate}
\item if $e_1, \ldots, e_D$ do not all intersect $p$ or
\item $\det \left( \frac{\partial x^a_{e_1, \ldots, e_D}}{\partial (t_1, \ldots, t_D)} \right)_p=0$ (which is a diffeomorphism invariant statement).
\end{enumerate}

Otherwise it tends to 
\be  \frac{1}{2^D} \text{sgn} \left( \det \left( \frac{\partial x^a_{e_1, \ldots, e_D}}{\partial (t_1, \ldots, t_D)} \right) \right)_p \hat{O}_{e_1, \ldots, e_D}(p) \prod_{i=1}^D \Delta^i_D \text{.} \ee 
Here we have denoted by $\hat{O}_{e_1, \ldots, e_D}(p)$ the trace(s) involved in the various terms of (\ref{eq:volf}). 

\end{lemma}

We conclude that (\ref{eq:volf}) reduces to 
\be  &\m& \lim_{\Delta_D \rightarrow 0} \hat{\pi}(p, \Delta_1, \ldots, \Delta_D) f \nonumber \\ &=& \sum_{e_1, \ldots, e_D} \frac{(i \hbar)^D s(e_1, \ldots, e_D)}{2^D D! \text{vol}(\Delta_1) \ldots \text{vol}(\Delta_{D-1}) } \chi_{\Delta_1}(p, v) \ldots \chi_{\Delta_{D-1}}(p, v ) \hat{O}_{e_1, \ldots, e_D}(0, \ldots, 0) \text{,}  \ee 
where $v$ on the right-hand side is the intersection point of the $D$-tuple of edges and it is understood that we only sum over such $D$-tuples of edges which are incident at a common vertex and $s(e_1, \ldots, e_D) := \text{sgn}(\det (\dot{e}_1(0), \ldots, \dot{e}_D(0)))$. Moreover, 
\be  \hat{O}_{e_1, \ldots, e_D}(0, \ldots, 0)  = \frac{1}{2} \epsilon_{IJI_1 J_1 I_2 J_2 \ldots I_nJ_n} R^{IJ}_e R^{I_1K_1}_{e_1} R^{J_1}_{e'_1} \m_{K_1}  \ldots R^{I_nK_n}_{e_n} R^{J_n}_{e'_n} \m_{K_n}   \ee 
and
\be R^{IJ}_e := R^{IJ}(h_e(0,1)) := \text{tr} \left( (\tau^{IJ} h_e(0,1))^T \frac{\partial}{\partial h_e(0,1)} \right)  \ee 
is a right-invariant vector field in the $\tau^{IJ}$ direction of SO$(D+1)$, that is, $R(hg) = R(h)$. We have also extended the values of the sign function to include $0$, which takes care of the possibility that one has $D$-tuples of edges with linearly dependent tangents. 

The final step is choosing $\Delta_1 = \ldots = \Delta_{D-1}$ and exponentiating the modulus by $1/(D-1)$. We replace the sum over all $D$-tuples incident at a common vertex $\sum_{e_1, \ldots, e_D}$ by a sum over all vertices followed by a sum over all  $D$-tuples incident at the same vertex $\sum_{v \in V(\gamma)} \sum_{e_1 \cap \ldots \cap e_D = v}$. Now, for small enough $\Delta$ and given $p$, at most one vertex contributes, that is, at most one of $\chi_{\Delta}(v,p) \neq 0$ because all vertices have finite separation. Then we can take the relevant $\chi_{\Delta}(p,v) = \chi_{\Delta}(p,v)^2$ out of the exponential and take the limit, which results in 

\begin{eqnarray}
 \hat{V}(R) &=& \int_R d^Dp \, \widehat{\left| \det (q)(p) \right| \m_\gamma }= \int_R d^Dp \hat{V}(p)_\gamma \text{,} \\
 \hat{V}(p) &=& \left( \frac{\hbar}{2} \right)^{\frac{D}{D-1}} \sum_{v \in V(\gamma)} \delta^D(p,v) \hat{V}_{v,\gamma} \text{,} \\
 \hat{V}_{v,\gamma} &=& \left| \frac{i^D}{D!} \sum_{e_1, \ldots, e_D \in E(\gamma), \, e_1 \cap \ldots \cap e_D = v} s(e_1, \ldots, e_D) q_{e_1, \ldots, e_D} \right|^{\frac{1}{D-1}} \text{,} \\
 q_{e_1, \ldots, e_D} &=& \frac{1}{2} \epsilon_{IJI_1 J_1 I_2 J_2 \ldots I_nJ_n} R^{IJ}_e R^{I_1K_1}_{e_1} R^{J_1}_{e'_1} \m_{K_1}  \ldots R^{I_nK_n}_{e_n} R^{J_n}_{e'_n} \m_{K_n} \text{.}
\end{eqnarray}

\subsubsection{$D+1$ Odd}

The case $D+1$ uneven works analogously, except that the expression for $\det(q)$ is changed a bit. With $n=D/2$, the result is

\begin{eqnarray}
 \hat{V}(R) &=& \int_R d^Dp \, \widehat{\left| \det (q)(p) \right| \m_\gamma }= \int_R d^Dp \hat{V}(p)_\gamma \text{,} \\
 \hat{V}(p) &=& \left( \frac{\hbar}{2} \right)^{\frac{D}{D-1}} \sum_{v \in V(\gamma)} \delta^D(p,v) \hat{V}_{v,\gamma} \text{,} \\
 \hat{V}^I_{v,\gamma} &=&  \frac{i^D}{D!} \sum_{e_1, \ldots, e_D \in E(\gamma), \, e_1 \cap \ldots \cap e_D = v} s(e_1, \ldots, e_D) q^I_{e_1, \ldots, e_D} \text{,}  \\
 \hat{V}_{v,\gamma} &=& \left|  \hat{V}^I_{v,\gamma}  \hat{V}_{I \, v,\gamma}  \right|^{\frac{1}{2D-2}}\text{,} \\
 q^I_{e_1, \ldots, e_D} &=&  \epsilon_{I} \m_{I_1 J_1 I_2 J_2 \ldots I_nJ_n}  R^{I_1K_1}_{e_1} R^{J_1}_{e'_1} \m_{K_1}  \ldots R^{I_nK_n}_{e_n} R^{J_n}_{e'_n} \m_{K_n} \text{.}
\end{eqnarray}

\subsubsection{More Results and Open Questions}

The derivations of cylindrical consistency, symmetry, positivity, self-adjointness and anomaly-freeness given in \cite{ThiemannModernCanonicalQuantum} generalise immediately to the higher dimensional volume operator. The question of uniqueness of the prefactor \cite{GieselConsistencyCheckOn1, GieselConsistencyCheckOn2}
in front of the expression under the square root of the volume operator  
or the computation of the matrix elements \cite{BrunnemannSimplificationOfThe, BrunnemannSimplificationOfThe, BrunnemannPropertiesOfThe1, BrunnemannPropertiesOfThe2, BrunnemannOrientedMatroidsCombinatorial} have not been addressed so far, however these are not necessary steps in order to use the volume operator for a consistent quantisation of the Hamiltonian constraint in what follows. We leave these open questions for 
future research.

\section{Implementation of the Hamiltonian Constraint}

\subsection{Introductory Remarks}

The implementation of the Hamiltonian constraint will follow along the lines of \cite{ThiemannModernCanonicalQuantum}, see \cite{ThiemannQSD1} for original literature and details. In our companion papers \cite{BTTI,BTTII}, we derived the classical expression 
\be  \tilde{\mathcal{H}} := \frac{\beta^2}{\sqrt{q}} \left( - \m^{(\beta)} \mathcal{H}_E + \frac{1}{2} \m^{(\beta)} D^{ab}_{\overline{M}} \; \left( \m^{(\beta)} F^{-1} \right) \m_{ab}^{\overline{M}} \, \m_{cd}^{\overline{N}} \; \m^{(\beta)} D^{cd}_{\overline{N}} - \left( \beta^2+1 \right)  K_{aI} K_{bJ} E^{a[I} E^{b|J]} \right) \text{,}  \ee 
where $a,b,c,\ldots$ are spatial indices and $I,J,K,\ldots$ are so$(D+1)$ indices. In order to have a well defined quantum version of this constraint, we have to express it in terms of holonomy and flux variables. As in the $3+1$-dimensional case, the volume operator turns out to be a cornerstone of the quantisation. 

At first, we will introduce a graph adapted triangulation of $\sigma$ in order to regularise the Hamiltonian constraint. Next, classical identities to express the Hamiltonian constraint in terms of holonomies and fluxes are derived. Since the complete expression for the Hamiltonian constraint will turn out to be rather laborious to write down, we will derive the regularisation piece by piece. Next, we show how to assemble the regularised pieces to the complete constraint and describe the quantisation. Finally, we construct a Hamiltonian Master constraint in order to avoid some of the usual difficulties associated with quantisation.

\subsection{Triangulation}

A natural choice for a triangulation turns out to be the following (we simplify the presentation drastically, the details can be found in \cite{ThiemannQSD1}): given a graph $\gamma$ one constructs a triangulation $T(\gamma, \epsilon)$ of $\sigma$ \em adapted \em to $\gamma$ which satisfies the following basic requirements.
\renewcommand{\labelenumi}{(\alph{enumi})}
\begin{enumerate}

\item The graph $\gamma$ is embedded in $T(\gamma, \epsilon)$ for all $\epsilon > 0$. 

\item The valence of each vertex $v$ of $\gamma$, viewed as a vertex of the infinite graph $T(\gamma, \epsilon)$, remains constant and is equal to the valence of $v$, viewed as a vertex of $\gamma$, for each $\epsilon > 0$. 

\item Choose a system of semianalytic\footnote{Semianalyticity is a more precise version
of piecewise analytic. See \cite{LewandowskiUniquenessOfDiffeomorphism} for complete definitions.} arcs $a^\epsilon_{\gamma,v,e,e'}$, one for each pair of edges $e,e'$ of $\gamma$ incident at a vertex $v$ of $\gamma$, which do not intersect $\gamma$ except in its endpoints where they intersect transversally. These endpoints are interior points of $e,e'$ and are those vertices of $T(\gamma, \epsilon)$ contained in $e,e'$ closest to $v$ for each $\epsilon > 0$ (i.e., no others are in between). For each $\epsilon, \epsilon ' > 0$ the arcs $a^\epsilon_{\gamma,v,e,e'}$, $a^{\epsilon '}_{\gamma,v,e,e'}$ are diffeomorphic with respect to semianalytic diffeomorphisms. The segments $e, e'$ incident at $v$ with outgoing orientation that are determined by the endpoints of the arc $a^\epsilon_{\gamma,v,e,e'}$ will be denoted by $s^\epsilon_{\gamma,v,e}$, $s^\epsilon_{\gamma,v,e'}$ respectively. Finally, if $\phi$ is a semianalytic diffeomorphism then $s^\epsilon_{\phi(\gamma),\phi(v),\phi(e)}$, $a^\epsilon_{\phi(\gamma),\phi(v),\phi(e),\phi(e')}$ and $\phi(s^\epsilon_{\gamma,v,e})$, $\phi(a^\epsilon_{\gamma,v,e,e'})$ are semianalytically diffeomorphic. 

\item Choose a system of mutually disjoint neighbourhoods $U^\epsilon_{\gamma, v}$, one for each vertex $v$ of $\gamma$, and require that for each $\epsilon > 0$ the $a^\epsilon_{\gamma,v,e,e'}$ are contained in $U^\epsilon_{\gamma, v}$. These neighbourhoods are nested in the sense that $U^\epsilon_{\gamma, v} \subset U^{\epsilon '}_{\gamma, v}$ if $\epsilon < \epsilon '$. and $\lim_{\epsilon \rightarrow 0} U^\epsilon_{\gamma, v} = \{v\}$. 

\item Triangulate $U^\epsilon_{\gamma, v}$ by $D$-simplices $\Delta(\gamma,v,e_1, \ldots, e_D)$, one for each ordered $D$-tuple of distinct edges $e_1, \ldots, e_D$ incident at $v$, bounded by the segments $s^\epsilon_{\gamma,v,e_1}, \ldots, s^\epsilon_{\gamma,v,e_D}$ and the arcs $a^\epsilon_{\gamma,v,e_1,e_2}, a^\epsilon_{\gamma,v,e_1,e_3}, \ldots, a^\epsilon_{\gamma,v,e_{D-1},e_D}$ ($D(D-1)/2$ arcs) from which loops $\alpha^\epsilon_{\gamma;v;e_1,e_2}$, etc. are built and triangulate the rest of $\sigma$ arbitrarily. The ordered $D$-tuple $e_1, \ldots, e_D$ is such that their tangents at $v$, in this sequence, form a matrix of positive determinant. 

\end{enumerate}

Requirement (a) prevents the action of the Hamiltonian constraint operator from being trivial.  Requirement (b) guarantees that the regulated operator $\hat{H}^{\epsilon}(N)$ is densely defined for each $\epsilon$.  Requirements (c), (d) and (e) specify the triangulation in the neighbourhood of each vertex of $\gamma$ and leave it unspecified outside of them. 

The reason why those $D$-simplices lying outside the neighbourhoods of the vertices described above are irrelevant will rest crucially on the choice of ordering with $[\hat{h}_s^{-1}, \hat{V}]$ on the rightmost: if $f$ is a cylindrical function over $\gamma$ and $s$ has support outside the neighbourhood of any vertex of $\gamma$, then $V(\gamma \cup s) - V(\gamma)$ consists of planar at most four-valent vertices only so that $[\hat{h}_s^{-1}, \hat{V}]f=0$.

We will define our operator on functions cylindrical over coloured graphs, that is, we define it on spin network functions. The domain for the operator that we will choose is a finite linear combination of spin-network functions, hence this defines the operator uniquely as a linear operator. Any operator automatically becomes consistent if one defines it on a basis, the consistency condition simply drops out.

The volume operator will appear in every term of the regulated Hamiltonian constraint. We will choose a factor ordering such that the Hamiltonian constraint acts only on vertices. It is therefore sufficient to regularise the constraint at vertices. As in the usual treatment, we use the tangents to the edges at a vertex as tangent vectors spanning the tangent space of the spatial coordinates. To emphasise this, we will abuse the notation in the following way: Let $e_a(\Delta)$ denote the $D$ edges incident at the vertex $v$ of an analytic $D$-simplex $\Delta \in T(\gamma, \epsilon)$. The matrix consisting of the tangents of the edges $e_1(\Delta), \ldots, e_D(\Delta)$ at $v$ (in that sequence) has non-negative determinant, which induces an orientation of $\Delta$. Furthermore, let $\alpha_{ab}$ be the arc on the boundary of $\Delta$ connecting the endpoints of $e_a(\Delta)$, $e_b(\Delta)$ such that the loop $\alpha_{ab}(\Delta) = e_a(\Delta) \circ a_{ab} (\Delta) \circ e_b(\Delta)^{-1}$ has positive orientation in the induced orientation of the boundary for $a<b$ (modulo cyclic permutation) and negative in the remaining cases. 

\subsection{Key Classical Identities}

The following classical identities are key for the rest of the discussion.

\subsubsection{$D+1 \geq 3$ Arbitrary} 

We observe that
\be  \sqrt{q} \pi_{aIJ}(x) := -(D-1) \{ A_{aIJ},  V(x, \epsilon)  \} \text{,} \ee 
where $V(x, \epsilon):=\int d^Dy \, \chi_\epsilon(x,y) \sqrt{q}$ is the volume of the region defined by $\chi_\epsilon(x,y)=1$ measured by $q_{ab}$ and $\chi_\epsilon(x,y) = \prod_{a=1}^D \Theta(\epsilon/2-|x^a-y^a|)$ is the characteristic function of a cube of coordinate volume $\epsilon^D$ with centre $x$.
Also, 
\be n^I(x) n_J(x)  \approx \frac{1}{D-1} \left( \pi^{aKI}(x) \pi_{aKJ}(x)- \eta^{I} \m_J \right) \text{.} \ee 

We can write the $KKEE$ terms in the same way as in the usual $3+1$-dimensional case, using 
\be  K(x) := E^{aI}(x) K_{aI}(x) \approx \frac{D-1}{D} \left\{  \mathcal{H}_E(x),   V(x, \epsilon) \right\} \text{.}  \ee 
Further,
\be  E^{bI}(x) K_{aI}(x) \approx \frac{(D-1)}{2D} \pi^{bKL}(x) \left\{A_{aKL}(x) , \left\{ \mathcal{H}_E[1](x, \epsilon), V(x, \epsilon) \right\} \right\}  \ee 
gives us access to all the needed terms.

\subsubsection{$D+1$ Even}
Let $n=(D-1)/2$. It is easy to see that
\begin{eqnarray}
   \pi^{aIJ}(x) & \approx & \frac{1}{(D-1)!} \epsilon^{ab_1c_1 \ldots b_n c_n} \epsilon^{IJI_1J_1 \ldots I_n J_n} \text{sgn} (\det e)(x) \nonumber \\
    & & \pi_{b_1 I_1 K_1}(x) \pi_{c_1 J_1} \m^{K_1}(x) \ldots \pi_{b_n I_n K_n}(x) \pi_{c_n J_n} \m^{K_n}(x) \sqrt{q}^{D-1}(x)  \text{.} 
\end{eqnarray}
The sign of the determinant of $e_a^I$ where the internal space is the subspace perpendicular to $n^I$ is accessible through
\begin{eqnarray}
\text{sgn} (\det(e_a^I))(x) & \approx & \frac{1}{2 D!} \epsilon^{IJI_1J_1 \ldots I_n J_n} \epsilon^{aa_1b_1 \ldots a_n b_n} \sqrt{q}^{D-1} \pi_{aIJ}(x) \nonumber \\
 & &  \pi_{a_1 I_1 K_1}(x) \pi_{b_1 J_1} \m^{K_1}(x) \ldots \pi_{a_n I_n K_n}(x) \pi_{b_n J_n} \m^{K_n}(x) \text{.} 
\end{eqnarray}
For the Euclidean part of the Hamiltonian constraint, we need
\begin{alignat}{3}
 \frac{\pi^{[a|IK} \pi^{b]J} \m_K}{\sqrt{q}}(x)  \approx & \frac{1}{4 (D-2)!} \epsilon^{abca_1b_1\ldots a_{n-1} b_{n-1}} \epsilon^{IJKLI_1J_1\ldots I_{n-1} J_{n-1}} \text{sgn} (\det e)(x) \\
   & \pi_{cKL}(x) \pi_{a_1 I_1 K_1}(x) \pi_{b_1 J_1} \m^{K_1}(x) \ldots \pi_{a_{n-1} I_{n-1} K_{n-1}}(x) \pi_{b_{n-1} J_{n-1}} \m^{K_{n-1}}(x) \sqrt{q}^{D-2}(x) \text{.} \nonumber 
\end{alignat}
Regarding quantisation, we have to choose a classical expression for $\frac{\pi^{[a|IK} \pi^{b]J} \m_K}{\sqrt{q}}(x)$. The above expression would be favourable by arguments of simplicity if it would not contain the additional factor of $\text{sgn} (\det(e_a^I))(x)$ which has to be accounted for. Therefore, we can equally well express the two factors of $\pi^{aIJ}$ separately and absorb the inverse square root into volume operators. 

\subsubsection{$D+1$ Odd}
Let $n=(D-2)/2$. With only minor modifications of the $D+1$ even case, we get
\begin{eqnarray}
   \pi^{aIJ}(x) & \approx & \frac{1}{(D-1)!} \epsilon^{abb_1c_1 \ldots b_n c_n} \epsilon^{IJKI_1J_1 \ldots I_n J_n} \text{sgn} (\det e)(x) \pi_{bLK}(x) n^L(x)\nonumber \\
    & & \pi_{b_1 I_1 K_1}(x) \pi_{c_1 J_1} \m^{K_1}(x) \ldots \pi_{b_n I_n K_n}(x) \pi_{c_n J_n} \m^{K_n}(x) \sqrt{q}^{D-1}(x) 
\end{eqnarray}
with
\begin{eqnarray}
  n^I(x) & \approx & \frac{1}{D!} \epsilon^{a_1 b_1 \ldots a_{n+1} b_{n+1}} \epsilon^{II_1J_1 \ldots I_{n+1} J_{n+1}} \text{sgn} (\det e)(x) \sqrt{q}^{D-1}(x) \nonumber \\
  & & \pi_{a_1 I_1 K_1}(x) \pi_{b_1 J_1} \m^{K_1}(x) \ldots \pi_{a_{n+1} I_{n+1} K_{n+1}}(x) \pi_{b_{n+1} J_{n+1}} \m^{K_{n+1}}(x) \text{.} 
\end{eqnarray}
For the Euclidean part of the Hamiltonian constraint, we need
\begin{eqnarray}
 \frac{\pi^{[a|IK} \pi^{b]J} \m_K}{\sqrt{q}} & \approx & \frac{1}{2 (D-2)!} \epsilon^{aba_1b_1\ldots a_{n} b_{n}} \epsilon^{IJKI_1J_1\ldots I_{n} J_{n}} \text{sgn} (\det e) \nonumber \\
  & & n_K \pi_{a_1 I_1 K_1} \pi_{b_1 J_1} \m^{K_1} \ldots \pi_{a_{n} I_{n} K_{n}} \pi_{b_{n} J_{n}} \m^{K_{n}} \sqrt{q}^{D-2}  
\end{eqnarray}
and observe that the factor of $\text{sgn} (\det(e_a^I))(x)$ is canceled by another such factor coming from $n^I$. The Euclidean part of the Hamiltonian constraint therefore has the same amount of complexity, measured by the ``number of involved operators'', in even and odd dimensions. 

\subsection{General Scheme}
\label{sec_generalscheme}

The basic idea of the regularisation of the Hamiltonian constraint operator is to approximate the constraint operator on the graph adapted triangulation and then to take the limit of an infinitely refined triangulation. For this procedure to work, it is mandatory that the constraint operator has a density weight of $+1$. A typical term of the classical Hamiltonian constraint (or any other operator one wants to regulate) will, after using the above classical identities, consist of 
\begin{itemize}
\item an integral $\int_\sigma d^D x$,
\item $n \in \mathbb{N}_0$ spatial $\epsilon$ symbols,
\item factors of $A_{aIJ}(x)$,
\item Poisson brackets involving a factor of $A_{aIJ}(x)$ as one of its two arguments as well as either the volume of a neighbourhood of $x$, the Euclidean part of the Hamiltonian constraint smeared  with unit lapse over a region containing $x$, or the Poisson bracket of the Euclidean part of the Hamiltonian constraint with the volume, smeared as before, as the other argument,
\item field strength tensors,
\item a factor of $\sqrt{q}^{1-n}$,
\item (covariant) derivatives.
\end{itemize}
Operators that are well defined on the kinematical Hilbert space are holonomies and the volume operator. We will show in the following that we can construct the Euclidean part of the Hamiltonian constraint operator, which gives us access to the remaining part of the constraint operator. As a start, it is therefore mandatory to write the Euclidean part of the Hamiltonian constraint in terms of holonomies and volume operators. We stress that we do not quantise the $\pi^{aIJ}$ as flux operators, which would also be possible. The reason is that the Hamiltonian constraint operator would not simplify significantly by using fluxes instead of derived flux operators. On the other hand, the appearance of fluxes only through volume operators can be seen as a certain simplification. Anyhow, different regularisations are possible and the discrimination between different regularisations has to be considered in the semiclassical limit. 

We begin with rewriting the integral. Given a $D$-tuple of edges $(e_1, \ldots, e_D)$ incident at $v$ with outgoing orientation consider the $D$-simplex $\Delta^{\epsilon} (\gamma, e_1, \ldots, e_D)$ bounded by the $D$ segments $s^\epsilon_{\gamma, v, e_1}$, \ldots, $s^\epsilon_{\gamma, v, e_D}$ incident at $v$ and the $D(D-1)/2$ arcs $a^\epsilon_{\gamma, v, e_a, e_b}$, $1 \leq a < b \leq D$. We now define the ``mirror images'' 
\begin{eqnarray}
  s^\epsilon_{\gamma, v, \bar{p}}(t) &:=& 2v-s^\epsilon_{\gamma, v, p}(t) \label{eq:mirrorimages} \text{,} \nonumber \\
  a^\epsilon_{\gamma, v, \bar{p}, \bar{p}'}(t) &:=& 2v-a^\epsilon_{\gamma, v, p, p'}(t) \text{,} \nonumber \\
  a^\epsilon_{\gamma, v, \bar{p}, p'}(t) &:=& a^\epsilon_{\gamma, v, \bar{p}, \bar{p}'}(t)-2t[v-s^\epsilon_{\gamma, v, p'}(1)] \text{,} \nonumber \\
  a^\epsilon_{\gamma, v, p, \bar{p}'}(t) &:=& a^\epsilon_{\gamma, v, p, p'}(t)+2t[v-s^\epsilon_{\gamma, v, p'}(1)] \text{,}
\end{eqnarray}  
where $p \neq p' \in {e_1, \ldots, e_D}$ and we have chosen some parametrisation of segments and arcs. Using the data (\ref{eq:mirrorimages}) we build $2^D-1$ more ``virtual'' $D$-simplices bounded by these quantities so that we obtain altogether $2^D$ $D$-simplices that saturate $v$ and triangulate a neighbourhood $U^\epsilon_{\gamma,v,e_1, \ldots, e_D}$ of $v$. Let $U^\epsilon_{\gamma,v}$ be the union of these neighbourhoods as we vary the ordered 
$D$-tuple of edges of $\gamma$ incident at $v$. The $U^\epsilon_{\gamma,v}$, $v \in V(\gamma)$ were chosen to be mutually disjoint in point (d) above. Let now 
\begin{eqnarray}
  \bar{U}^\epsilon_{\gamma,v,e_1, \ldots, e_D} &:=& U^\epsilon_{\gamma,v} - U^\epsilon_{\gamma,v,e_1, \ldots, e_D} \text{,} \nonumber \\
  \bar{U}^\epsilon_{\gamma} &:=& \sigma - \bigcup_{v \in V(\gamma)} U^\epsilon_{\gamma,v} \text{,}
\end{eqnarray}
then we may write any classical integral (symbolically) as
\begin{eqnarray}
  \int_\sigma &=& \int_{\bar{U}^\epsilon_\gamma} + \sum_{v \in V(\gamma)} \int_{U^\epsilon_{\gamma,v}} \label{eq:integraldecomposition} \nonumber \\
  &=& \int_{\bar{U}^\epsilon_\gamma} + \sum_{v \in V(\gamma)} \frac{1}{E(v)} \sum_{v=b(e_1) \cap \ldots \cap b(e_D)} \left( \int_{U^\epsilon_{\gamma,v,e_1, \ldots, e_D}} + \int_{\bar{U}^\epsilon_{\gamma,v,e_1, \ldots, e_D}}\right) \nonumber \\
  &\approx& \int_{\bar{U}^\epsilon_\gamma} + \sum_{v \in V(\gamma)} \frac{1}{E(v)} \left[ \sum_{v=b(e_1) \cap \ldots \cap b(e_D)} 2^D \int_{\Delta^\epsilon_{\gamma,v,e_1, \ldots, e_D}} + \int_{\bar{U}^\epsilon_{\gamma,v,e_1, \ldots, e_D}}\right] \text{,}
\end{eqnarray}
where in the last step we have noticed that classically the integral over $U^\epsilon_{\gamma,v,e_1, \ldots, e_D}$ converges to $2^D$ times the integral over $\Delta^\epsilon_{\gamma,v,e_1, \ldots, e_D}$, $\approx$ means approximately and $E(v) = \binom{n(v)}{D}$ with $n(v)$ being the valence of the vertex. Now when triangulating the regions of the integrals over $\bar{U}^\epsilon_{\gamma,v,e_1, \ldots, e_D}$ and $\bar{U}^\epsilon_{\gamma}$ in (\ref{eq:integraldecomposition}), regularisation and quantisation gives operators that vanish on $f_\gamma$ because the corresponding regions do not contain a non-planar vertex of $\gamma$.

As a next step, we approximate the integral 
\be  \int_{\Delta^\epsilon_{\gamma,v,e_1, \ldots, e_D}} d^Dx \,g(x) \approx \frac{1}{D!} \epsilon^D g(v)  \ee 
for some function $g(x)$. Here we assumed the coordinate length of each segment $s^\epsilon_{\gamma, v, e_a}$ to be $\epsilon$. The general case of arbitrary coordinate length works analogously, since the factors of $\epsilon$ will be hidden in holonomies and derivatives contracted with an epsilon symbol which addresses each segment exactly once. The factor $1/D!$ accounts for the volume of a $D$-simplex. We now multiply the nominator and the denominator by $\epsilon^{D(n-1)}$. Together with the factors $\sqrt{q}^{1-n}(v)$ and the factor $\epsilon^D$ from the integral, we get $\epsilon^{Dn} / V(v,\epsilon)^{n-1}$. The volumes in the denominator are absorbed into the Poisson brackets by the standard technique. The factors of $A_{aIJ}$ are turned into holonomies $ (h_{s_a})_{KL} = \delta_{KL} + \epsilon \dot{e}^a(0) A_{aIJ} \left(\tau^{IJ}\right)_{KL} + \mathcal{O}(\epsilon^2)$ using the the same amount of factors of $\epsilon$ since we note that the zeroth order of the expansion of the holonomies vanishes when inserted into the Poisson brackets. We abbreviated $s_a = s^\epsilon_{\gamma, v, e_a}$ to simplify notation.

The field strength tensors can be dealt with as follows. Let $e, e'$ be arbitrary paths which are images of the interval $[0,1]$ under the corresponding embeddings, which we also denote by $e,e'$ such that $v=e(0)=e'(0)$. For any $0 < \epsilon < 1$ set $e_\epsilon(t) := e(\epsilon t)$ for $t \in [0,1]$ and likewise for $e'$. Then we expand $h_{e_\epsilon}(A)$ in powers of $\epsilon$. Consider the loop $\alpha_{e_\epsilon, e'_\epsilon}$ where in a coordinate neighbourhood
\be  
\alpha_{e_\epsilon, e'_\epsilon}(t)=\begin{cases}

  e_\epsilon(4t)  & 0 \leq t \leq 1/4 \\
  e_\epsilon(1)+e'_\epsilon(4t-1)-v  & 1/4 \leq t \leq 1/2 \\
  e'_\epsilon(1)+e_\epsilon(3-4t)-v  & 1/2 \leq t \leq 3/4 \\
  e'_\epsilon(4-4t)  & 3/4 \leq t \leq 1 \text{.}
\end{cases} 
 \ee 
Now expanding again in powers of $\epsilon$ we easily find $h_{\alpha_{e_\epsilon, e'_\epsilon}} = 1_{D+1} + \epsilon^2 F_{abIJ} \tau^{IJ} \dot{e}^a(0) \dot{e}'^b(0) + \mathcal{O}(\epsilon^3)$. Since the indices of the field strength tensors are contracted only with other antisymmetric index pairs, the zeroth order of the expansion vanishes as well as the orders beyond $\epsilon^2$ in the limit $\epsilon \rightarrow 0$.
The remaining factors of $\epsilon$ are absorbed into covariant derivatives using the approximation
\begin{eqnarray}
& & \left(h_e(0,\epsilon) \pi^a (e(\epsilon)) h_e(0, \epsilon)^{-1} -\pi^a(v) \right)^{AB} \nonumber \\
&=& \left((1+\epsilon \dot{e}^b(0) A_b) (\pi^b (v)+\epsilon \dot{e}^c(0) \partial_c \pi^b(v)) (1-\epsilon \dot{e}^d(0) A_d) -\pi^b(v) \right)^{AB} + \mathcal{O}(\epsilon^2) \nonumber  \\
&=&\epsilon \dot{e}^c(0) D_c \pi^{aAB}(v) + \mathcal{O}(\epsilon^2) \text{.}
\end{eqnarray}
We note that partial derivatives can be dealt with in the same way. 

At this point, all factors of $\epsilon$ have been absorbed into holonomies and derivatives. It is key that the volume operators are ordered to the right in the quantum theory since then, the Hamiltonian constraint evaluated on a cylindrical function $f_\gamma$ will only act on the vertices of $\gamma$. The action at vertices however does not depend on the value of $\epsilon > 0$ and we can take the limit $\epsilon \rightarrow 0$, thus removing the regulator. 

In order to quantise the Hamiltonian constraint, we have to replace the holonomies by multiplication operators, the volumes by volume operators, and the Poisson brackets by $i/\hbar$ times the commutator.

\subsection{Regularised Quantities}

\label{sec:RegularisedQuantities}

In order to construct a well defined Hamiltonian constraint operator, we have to express it in terms of operators well defined on the kinematical Hilbert space. Instead of writing down the explicit regularisation for the proposed Hamiltonian constraint, we want to provide a toolkit  for a general class of operators. In the following, we will propose ``regulated'' versions of the phase space variables, marked by an upper $\m^\epsilon$ in front. The idea will be to replace all phase space variables in the classical Hamiltonian constraint by their corresponding regulated versions, do some additional minor modifications and directly arrive at the Hamiltonian constraint operator, without explicitly dealing with the triangulation and the correct powers of $\epsilon$. Since the final constraint operator will only act on vertices of $\gamma$, it is sufficient to regularise the phase space variables at vertices $v$. 

In what follows, we use a graph adapted coordinate system, meaning that the spatial coordinates $a, b, \ldots = 1, \ldots, D$ enumerate the $D$ edges incident at $v$ of a $D$-simplex. 

\subsubsection{$D+1 \geq 3$ Arbitrary}

We will express all the basic variables in terms of holonomies living on the edges of the adapted triangulation and volume operators acting on it. First, we notice that 
\be  \m^\epsilon  (\sqrt{q}^{x+1} \pi_{aIJ}(v)) := \frac{(D-1)}{(x+1)} (h_{s_a})_{I} \m^K \{ (h_{s_a})^{-1}_{KJ}, \left( V(v, \epsilon) \right)^{x+1} \}  \ee 
is gauge covariant and reduces to $\epsilon \sqrt{q}^{x+1} \pi_{aIJ}(v)$ in the limit $\epsilon \rightarrow 0$. The factor of $\epsilon$ is expected as the regulated quantity has a lower spatial index. In the end, when the complete constraint operator will be assembled, all factors of $\epsilon$ will cancel out. We restrict $x > -1$ because powers of the volume operator will be defined by the spectral theorem in the quantum theory.

For the $KKEE$ terms, we propose
\begin{alignat}{3}
  \m^\epsilon \left( \frac{1}{\sqrt{q}} K_{aI} K_{bJ} E^{[a|I} E^{b]J} \right)  \approx  \frac{(D-1)^2}{4D^2} &   \m^\epsilon (\sqrt[4]{q}^{-1} \pi^{[a|KL}(v)) (h_{s_a})_{K} \m^O \left\{(h_{e_a})^{-1}_{OL} , \left\{  \mathcal{H}_E[1](v, \epsilon), V(v, \epsilon) \right\} \right\} \nonumber \\
 \times \; &   \m^\epsilon (\sqrt[4]{q}^{-1} \pi^{b]MN}(v)) (h_{s_b})_{M} \m^P \left\{(h_{e_b})^{-1}_{PN} , \left\{  \mathcal{H}_E[1](v, \epsilon), V(v, \epsilon) \right\} \right\} \nonumber \text{,} \\
\end{alignat}
where the $\m^\epsilon \pi^{aIJ}$ will be defined below. 

Next, we regulate the gauge unfixing term $DF^{-1}D$ with density weight 1. We will place zero density into $F^{-1}$ and a density weight of $1/2$ into each $D$. Accordingly, 
\be  \sqrt{q}^4 \left( F^{-1} \right) \m_{cd,}^{\overline{N}} \m_{ab}^{\overline{M}} = \gamma \sqrt{q}^4 \epsilon^{EFGH \overline{N}} \pi_{(c|EF} \left(F^{-1}\right)_{d)GH,(a|AB} \pi_{b)CD} \epsilon^{ABCD\overline{M}}  \ee 
becomes
\be    \m^\epsilon \left(\sqrt{q}^4 F^{-1} \right) \m_{cd,}^{\overline{N}} \m_{ab}^{\overline{M}} = \gamma \epsilon^{EFGH \overline{N}} \m^\epsilon(\sqrt{q} \pi_{(c|EF}) \m^\epsilon \left(\sqrt{q}^2 F^{-1}\right)_{d)GH,(a|AB} \m^\epsilon(\sqrt{q} \pi_{b)CD}) \epsilon^{ABCD\overline{M}}  \ee 
with
\begin{eqnarray}
 \m^\epsilon \left(\sqrt{q}^2 F^{-1} \right)_{aIJ,bKL} &:=& \frac{1}{4 (D-1)} \m^\epsilon(\sqrt{q} \pi_{aAC}) \m^\epsilon(\sqrt{q} \pi_{bBD}) \left(\m^\epsilon(\sqrt{q}^{-1} \pi^{cEC}) \m^\epsilon(\sqrt{q}  \pi_{cE} \m^D) - \eta^{CD} \right) \nonumber \\
 & & \left( \eta^{AB} \eta^{K[I} \eta^{J]L} - 2 \eta^{LA} \eta^{B[I} \eta^{J]K} \right) \text{.} 
\end{eqnarray}
The $D$ constraint contains a covariant derivative which we regularise as
\begin{eqnarray}
 \m^\epsilon (\sqrt{q}^{-1} D_a \pi^{bAB}) := \left( h_{s_a}  \m^\epsilon (\sqrt{q}^{-1} \pi^b (s_a)) h_{s_a}^{-1} -\m^\epsilon (\sqrt{q}^{-1} \pi^b(v)) \right)^{AB} \text{.}
\end{eqnarray}
The full $D$ constraint 
\be D^{ab}_{\overline{M}} = -\epsilon_{IJKL\overline{M}} \pi^{cIJ} \left( \pi^{(a|KN} D_c \pi^{b)L} \m_N \right) \ee 
can thus be regularised as
\begin{eqnarray}
 \m^\epsilon (\sqrt{q}^{-3/2} D^{ab}_{\overline{M}}) = -\epsilon_{IJKL\overline{M}} \m^\epsilon (\sqrt{q}^{-1/2} \pi^{cIJ}) \left( \m^\epsilon (\sqrt{q}^{-1} \pi^{(a|KN}) \m^\epsilon (\sqrt{q}^{-1} D_c \pi^{b)L}\m_N) \right) \text{.}
\end{eqnarray}
A different regularisation procedure for the $D F^{-1} D$ part of the Hamiltonian constraint which is based on field strength tensors is outlined in appendix \ref{app:Regularisierung}.

In general, a generic power of $1/\sqrt{q}$ needed to turn the individual terms with densities $>1$ into densities of weight $1$ can be constructed as
\be  \m^\epsilon  \left( \frac{1}{\sqrt{q}^{(-2xD-2)}} \right) \approx \left( \frac{1}{2} \right)^D \det \left( \m^\epsilon  (\sqrt{q}^{x+1} \pi_{aIJ}) \m^\epsilon  (\sqrt{q}^{x+1} \pi_b \m^{IJ}) \right)  \ee 
with the usual $x>-1$.

The field strength tensors are regularised as 
\be \m^\epsilon F_{abIJ}   = \left(h_{\alpha_{s_a, s_b}}\right)_{KL} \delta_{[I}^K \delta_{J]}^L  \ee 
while we set
\be  \m^\epsilon \{A_{aIJ}(v), \cdot\} = - (h_{s_a})_{I} \m^K \{  (h_{s_a} \m^{-1})_{KJ}, \cdot \} \text{.}  \ee 

\subsubsection{$D+1$ Even}

Let $n=(D-1)/2$. We ``regulate''
\begin{eqnarray}
   \m^\epsilon (\sqrt{q}^{(D-1)x} \pi^{aIJ}(v)) & \approx & \frac{1}{(D-1)!} \epsilon^{ab_1c_1 \ldots b_n c_n} \epsilon^{IJI_1J_1 \ldots I_n J_n} \text{sgn} (\det e)(v) \nonumber \\
    & & \m^\epsilon (\sqrt{q}^{(1+x)} \pi_{b_1 I_1 K_1}(v)) \m^\epsilon (\sqrt{q}^{(1+x)} \pi_{c_1 J_1} \m^{K_1}(v)) \ldots \nonumber \\
    & & \m^\epsilon (\sqrt{q}^{(1+x)} \pi_{b_n I_n K_n}(v)) \m^\epsilon (\sqrt{q}^{(1+x)} \pi_{c_n J_n} \m^{K_n}(v)) 
\end{eqnarray}
and
\begin{eqnarray}
\m^\epsilon( \text{sgn} (\det(e_a^I))) & \approx & \frac{1}{2 D!} \epsilon^{IJI_1J_1 \ldots I_n J_n} \epsilon^{aa_1b_1 \ldots a_n b_n}  \m^\epsilon (\sqrt{q}^{(D-1)/D} \pi_{aIJ} ) \nonumber \\
 & &  \m^\epsilon (\sqrt{q}^{(D-1)/D} \pi_{a_1 I_1 K_1}) \m^\epsilon (\sqrt{q}^{(D-1)/D} \pi_{b_1 J_1} \m^{K_1}) \ldots \nonumber \\
 & &  \m^\epsilon (\sqrt{q}^{(D-1)/D} \pi_{a_n I_n K_n}) \m^\epsilon (\sqrt{q}^{(D-1)/D} \pi_{b_n J_n} \m^{K_n} ) \text{.} 
\end{eqnarray}
For the Euclidean part of the Hamiltonian constraint, we need
\begin{alignat}{3}
 \m^\epsilon \left( \frac{\pi^{[a|IK} \pi^{b]J} \m_K}{\sqrt{q}} \right)  \approx & \frac{1}{4 (D-2)!} \epsilon^{abca_1b_1\ldots a_{n-1} b_{n-1}} \epsilon^{IJKLI_1J_1\ldots I_{n-1} J_{n-1}} \text{sgn} (\det e)  \nonumber \\
  &  \m^{\epsilon}(\sqrt{q} \pi_{cKL}) \m^{\epsilon}(\sqrt{q} \pi_{a_1 I_1 K_1}) \m^{\epsilon}(\sqrt{q} \pi_{b_1 J_1} \m^{K_1}) \ldots \nonumber \\ 
  & \m^{\epsilon}(\sqrt{q}\pi_{a_{n-1} I_{n-1} K_{n-1}})  \m^{\epsilon}(\sqrt{q} \pi_{b_{n-1} J_{n-1}} \m^{K_{n-1}}) \text{.} ~~~~
\end{alignat}
As stressed before, the two possibilities to express the Euclidean part of the Hamiltonian constraint are equally complicated.

\subsubsection{$D+1$ Odd}
Let $n=(D-2)/2$. We ``regulate''
\begin{eqnarray}
   \m^\epsilon (\sqrt{q}^{(D-1)x} \pi^{aIJ}(v)) & \approx &  \frac{1}{(D-1)!} \epsilon^{abb_1c_1 \ldots b_n c_n} \epsilon^{IJKI_1J_1 \ldots I_n J_n} \text{sgn} (\det e)(v) \m^\epsilon (\sqrt{q}^{(1+x)}\pi_{bLK}(v)) \m^\epsilon n^L(v) \nonumber \\
    & & \m^\epsilon (\sqrt{q}^{(1+x)} \pi_{b_1 I_1 K_1}(v)) \m^\epsilon (\sqrt{q}^{(1+x)} \pi_{c_1 J_1}) \m^{K_1}(v) \ldots \nonumber \\
    & & \m^\epsilon (\sqrt{q}^{(1+x)} \pi_{b_n I_n K_n}(v)) \m^\epsilon (\sqrt{q}^{(1+x)} \pi_{c_n J_n} \m^{K_n}(v) ) 
\end{eqnarray}
and
\begin{eqnarray}
  \m^\epsilon n^I(v) & \approx & \frac{1}{D!} \epsilon^{a_1 b_1 \ldots a_{n+1} b_{n+1}} \epsilon^{II_1J_1 \ldots I_{n+1} J_{n+1}} \text{sgn} (\det e)(v) \nonumber \\
  & & \m^\epsilon (\sqrt{q}^{(D-1)/D} \pi_{a_1 I_1 K_1}(v)) \m^\epsilon (\sqrt{q}^{(D-1)/D} \pi_{b_1 J_1} \m^{K_1}(v)) \ldots \nonumber \\
  & & \m^\epsilon (\sqrt{q}^{(D-1)/D} \pi_{a_{n+1} I_{n+1} K_{n+1}}(v)) \m^\epsilon (\sqrt{q}^{(D-1)/D} \pi_{b_{n+1} J_{n+1}} \m^{K_{n+1}}(v))   \text{.}
\end{eqnarray}
For the Euclidean part of the Hamiltonian constraint, we need
\begin{eqnarray}
 \m^\epsilon \left( \frac{\pi^{[a|IK} \pi^{b]J} \m_K}{\sqrt{q}} \right) & \approx & \frac{1}{2 (D-2)!} \epsilon^{aba_1b_1\ldots a_{n} b_{n}} \epsilon^{IJKI_1J_1\ldots I_{n} J_{n}} \text{sgn} (\det e)  \\
  & & \m^\epsilon (n_K) \m^\epsilon (\sqrt{q} \pi_{a_1 I_1 K_1}) \m^\epsilon (\sqrt{q} \pi_{b_1 J_1} \m^{K_1}) \ldots \m^\epsilon (\sqrt{q}\pi_{a_{n} I_{n} K_{n}}) \m^\epsilon (\sqrt{q} \pi_{b_{n} J_{n}} \m^{K_{n}}) \nonumber \text{.}
\end{eqnarray}

\subsection{The Hamiltonian Constraint Operator}
At this point, we are ready to assemble the Hamiltonian constraint operator. The general idea of the regularisation has been described in section \ref{sec_generalscheme}. Here, we provide a 
toolkit in order to assemble the constraint operator. 
\renewcommand{\labelenumi}{(\arabic{enumi})}
\begin{enumerate}
  \item The Euclidean part $\mathcal{H}_E = \frac{1}{2\sqrt{q}} \pi^{aIK} \pi^{bJ} \m_K  F_{abIJ} $ of the Hamiltonian constraint can be quantised with the methods described above and using the following recipe. The corresponding operator can then be used in commutators to express additional parts of the full Hamiltonian constraint operator.   
  \item Use classical identities in order to express the Hamiltonian constraint in terms of connections $A_{aIJ}$, volumes $V(x, \epsilon)$ and Euclidean Hamiltonian constraints $\mathcal{H}_E(x, \epsilon)$. 
  \item Replace all phase space variables by their corresponding regulated quantities.
  \item Instead of the the integration $\int_\sigma d^Dx$, put a sum $\frac{1}{D!} \sum_{v \in V(\gamma)}$ over all the vertices $v$ of the graph $\gamma$. 
  \item For every spatial $\epsilon$-symbol, put a sum $\frac{2^D}{E(v)} \sum_{v(\Delta)=v}$ over all $D$-simplices having $v$ as a vertex. The holonomies associated with the $\epsilon$-symbol are evaluated along the edges spanning $\Delta$.
  \item Substitute the Poisson brackets by $\frac{i}{\hbar}$ times the commutator of the corresponding operators, i.e. the multiplication operator $\hat{h}_e$ and the volume operator $\hat{V}$.  
\end{enumerate}  

In order to understand the double sum over $D$-simplices appearing in the $KKEE$ and the gauge unfixing term, consider the following argument given in a similar form in \cite{ThiemannQSD5}: Since $\lim_{\epsilon \rightarrow 0} (1/\epsilon^D) \chi_\epsilon(x,y) = \delta^D(x,y)$ we have $\lim_{\epsilon \rightarrow 0} (1/\epsilon^D)V(x,\epsilon) = \sqrt{q}(x)$. It is also easy to see that for each $\epsilon > 0$ we have that $\delta V / \delta \pi^{aIJ}(x) = \delta V(x,\epsilon) / \delta \pi^{aIJ}(x)$. The terms under consideration are of the form
\be  \int d^Dx\, \frac{\sqrt{q}(x) \pi_{aIJ}(x) Z^{aIJ}(x) \sqrt{q}(x) \pi_{bKL}(x) Z^{bKL}(x)}{\sqrt{q}(x)} \text{,} \ee 
where $Z^{aIJ}$ is a density of weight $+1$ and stands symbolically for the remaining terms, including a spatial $\epsilon$-symbol with upper indices, one of which is $a$. We rewrite this expression as
\begin{alignat}{3}
   &\lim_{\epsilon \rightarrow 0} \frac{1}{\epsilon^D} 4(D-1)^2 \int d^Dx \, \frac{ \{A_{aIJ}(x), V\} Z^{aIJ}(x)}{2 \sqrt[4]{q}(x)} \int d^Dy\, \chi_\epsilon(x,y) \frac{ \{A_{bKL}(y), V\} Z^{bKL}(y)}{2 \sqrt[4]{q}(y)} \\
  =& \lim_{\epsilon \rightarrow 0} \frac{1}{\epsilon^D} 4(D-1)^2 \int d^Dx \, \frac{ \{A_{aIJ}(x), V(x, \epsilon)\} Z^{aIJ}(x)}{2 \sqrt[4]{q}(x)} \int d^Dy\, \chi_\epsilon(x,y) \frac{ \{A_{bKL}(y), V(y, \epsilon)\} Z^{bKL}(y)}{2 \sqrt[4]{q}(y)} \nonumber \\
  =& \lim_{\epsilon \rightarrow 0} \frac{1}{\epsilon^D} 4(D-1)^2 \int d^Dx \, \frac{ \{A_{aIJ}(x), V(x, \epsilon)\} Z^{aIJ}(x)}{2 \sqrt{V(y, \epsilon)/\epsilon^D}} \int d^Dy\, \chi_\epsilon(x,y) \frac{ \{A_{bKL}(y), V(y, \epsilon)\} Z^{bKL}(y)}{2 \sqrt{V(y, \epsilon)/\epsilon^D}} \nonumber \\
  =& \lim_{\epsilon \rightarrow 0}  4(D-1)^2 \int d^Dx \, \frac{ \{A_{aIJ}(x), V(x, \epsilon)\} Z^{aIJ}(x)}{2 \sqrt{V(y, \epsilon)}} \int d^Dy\, \chi_\epsilon(x,y) \frac{ \{A_{bKL}(y), V(y, \epsilon)\} Z^{bKL}(y)}{2 \sqrt{V(y, \epsilon)}} \nonumber \\
  =& \lim_{\epsilon \rightarrow 0}  4(D-1)^2 \int d^Dx \, \{A_{aIJ}(x), \sqrt{V(x, \epsilon)}\} Z^{aIJ}(x) \int d^Dy\, \chi_\epsilon(x,y)  \{A_{bKL}(y), \sqrt{V(y, \epsilon)}\} Z^{bKL}(y) \nonumber \text{.}
\end{alignat}  
Triangulation leads to two sums over vertices and two sums over $D$-simplices containing the individual vertices. In the limit $\epsilon \rightarrow 0$ however the two sums over vertices collapse to a single sum over vertices due to the $\chi_\epsilon$ term and we have the desired result.

\subsection{Solution of the Hamiltonian Constraint} 

As in the $3+1$-dimensional treatment, we realise that the only spin changing operation of the Hamiltonian constraint is performed by its Euclidean part. The construction of a set of rigorously defined solutions to the diffeomorphism and the Hamiltonian constraint described in \cite{ThiemannQSD2} thus immediately generalises to our case.

\subsection{Master Constraint}

The implementation of the Master constraint 
\be  \boldsymbol{M}  = \frac{1}{2} \int_\sigma d^Dx \, \frac{\mathcal{H}(x)^2}{\sqrt{q}(x)}  \ee 
works analogously to the $3+1$-dimensional case described in \cite{ThiemannQSD8}. The inverse square root is split up between the two Hamiltonian constraints and hidden by adjusting the power of the volume operators as before. The result of the derivation is the Master constraint operator
\be  \boldsymbol{\hat{M}} T_{[s]} := \sum_{[s_1]} Q_{\boldsymbol{M}}(T_{[s_1]}, T_{[s]}) T_{[s_1]}  \ee 
with
\be  Q_{\boldsymbol{M}}(l, l') = \sum_{[s]} \eta_{[s]} \sum_{v \in V(\gamma(s_0[s]))} \overline{l(\hat{C}^\dagger_v T_{s_0([s])})} l'(\hat{C}^\dagger_v T_{s_0([s])})  \ee 
and $l(\hat{C}^\dagger_v T_{s_0([s])})$ being the evaluation of $l$ on the Hamiltonian constraint operator with the additional $1/\sqrt[4]{q}$ hidden in the volume operator(s). The proof of the following theorem generalises with obvious modifications from the treatment in \cite{ThiemannModernCanonicalQuantum}.
\begin{theorem} \mbox{}\\
\renewcommand{\labelenumi}{(\roman{enumi})}
\begin{enumerate}
\item The positive quadratic form $Q_{\boldsymbol{M}}$ is closable and induces a unique, positive self-adjoint operator $\boldsymbol{\hat{M}}$ on $\mathcal{H}_{\text{diff}}$. 
\item Moreover, the point zero is contained in the point spectrum of $\boldsymbol{\hat{M}}$. 
\end{enumerate}
\end{theorem}
We deal with the problem of $\mathcal{H}_{\text{diff}}$ not being separable by using $\theta$-equivalence classes of spin-networks, see \cite{ThiemannQSD8}. Now, a direct integral decomposition of $\mathcal{H}^\theta_{\text{diff}}$ is available:
\begin{theorem} \m \\
There is a unitary operator $V$ such that $V \mathcal{H}^\theta_{\text{diff}}$ is the direct integral Hilbert space
\be  \mathcal{H}^\theta_{\text{diff}} \propto \int_{\mathbb{R}^+}^\oplus d\mu(\lambda) \, \mathcal{H}^\theta_{\text{diff}} (\lambda)  \ee 
where the measure class of $\mu$ and the Hilbert space $\mathcal{H}^\theta_{\text{diff}}(\lambda)$, in which $V \boldsymbol{\hat{M}} V^{-1}$ acts by multiplication by $\lambda$, are uniquely determined. 

The physical Hilbert space is given by $\mathcal{H}^\theta_{\text{phys}} = \mathcal{H}^\theta_{\text{diff}}(0)$.
\end{theorem}
We notice that we could define an extended Master Constraint that also involves the simplicity
constraint.

\subsection{Factor Ordering}

In \cite{GieselConsistencyCheckOn1, GieselConsistencyCheckOn2}, it has been shown that there is a unique factor ordering which results in a non-vanishing flux operator expressed through the volume operator and holonomies in the usual $3+1$ dimensional LQG. The idea, translated to our case, is that the volume operator in the expression for $\m^\epsilon \pi^{aIJ}$ has to act on an at least $D$-valent non-planar vertex and the holonomies in the expression have to be ordered to the right for this to be ensured. Apart from ordering individual terms of the sums appearing differently (which would be highly unnatural), this leaves only one possible factor ordering. We remark that the proof of the equivalence of the ``normal'' and ``derived'' flux operator given in \cite{GieselConsistencyCheckOn1, GieselConsistencyCheckOn2} does not generalise trivially to our case since it is explicitly based on SU$(2)$ as the internal gauge group. We leave this point open for further research. 

In order to ensure that the Hamiltonian constraint only acts on vertices, we order in all three terms either a commutator $[\hat{h}_e^{-1},\hat{V}]$ or a double-commutator $[\hat{h}_e^{-1},[\mathcal{H}_E,\hat{V}]$ to the right. 

We leave the remaining details of the factor ordering open, as this paper only intends to show that a quantisation is possible in principle.

\subsection{Outlook on Consistency Checks}

At this point, one might ask if there are good indications whether the proposed theory is physically viable. In case of the usual formulation of LQG in terms of Ashtekar-Barbero variables, it was shown in \cite{ThiemannQSD4} that a quantisation of Euclidean General Relativity in three dimensions with methods very similar to the ones used in LQG recovers the known solutions of three-dimensional General Relativity familiar from other approaches. The reason why these theories match is that they both use the gauge groups SU$(2)$ and that a suitable redefinition of the Lagrange multipliers of Euclidean three-dimensional General Relativity leads to a Hamiltonian constraint with the same algebraic structure as the Euclidean part of the constraint familiar from LQG. A similar check is conceivable for the presented theory in that we can describe Lorentzian three-dimensional General Relativity using SU$(2)$ as a gauge group, which would result in a different Hamiltonian constraint. One could now check if the solution space of Lorentzian three-dimensional General Relativity is reproduced when using SU$(2)$ as a gauge group and thus mimicking the internal signature switch which is also done in this formulation.

As for the simplicity constraint, we cannot use three-dimensional General Relativity as a testbed since the simplicity constraints only appear in four and higher dimensions. In this paper, two different regularisations for the gauge unfixing part of the Hamiltonian constraint were introduced, one in section \ref{sec:RegularisedQuantities} and one in appendix \ref{app:Regularisierung}. While the regularisation introduced in section \ref{sec:RegularisedQuantities} preserves the closure of the quantum constraint algebra, this is not obvious for the regularisation in appendix \ref{app:Regularisierung} since terms quadratic in the field strength appear.

Another approach to consistency checks is to compare our formulation in four dimensions to the usual LQG formulation. In section \ref{sec:Area}, the area operator was shown to have the same spectrum as in standard LQG, which however does not come as a surprise regarding similar results from spin foam models. As for the volume operator, we do not know whether the spectrum matches the one of standard LQG. This is also tied to the fact that we are only interested in the spectrum on the solution space to the vertex simplicity constraint operators, for which we do not have a completely satisfactory proposal. We remark that a matching spectrum of the volume operator can be obtained by using a weak implementation of the linear vertex simplicity constraints \cite{DingTheVolumeOperator}. However, as explained in our companion paper \cite{BTTV}, this approach comes with its own problems in the canonical theory.

\section{Conclusion}

In this paper we have demonstrated that by a straightforward adaption of the toolbox developed 
for LQG in $3+1$ dimensions also the constraints of our new connection formulation of General 
Relativity in any dimension $D+1\ge 3$ can be quantised analogously and rigorously. The 
higher dimension does not require much more complexity than in $3+1$ dimensions. We conclude 
that our new connection formulation has a consistent quantisation. The next task is to study matter coupling, in particular coupling to supersymmetric matter in interesting dimensions, where String theories and Supergravity theories are defined, and the quantisation thereof. This has to be done, as in $3+1$ dimensions, in a 
background independent way, a task to which we turn in the next papers of this series \cite{BTTIV, BTTVI, BTTVII}.

In four dimensions, we now have the special situation that there are two formulations of LQG, one based on the usual Asthekar-Barbero variables, and one based on the variables proposed in this series of papers. From a direct comparison, one concludes that the new formulation is more complicated since the Hamiltonian constraint contains an additional term resulting from gauge unfixing. Two different regularisations for this term were introduced, the first one directly regularises the covariant derivatives in this term, the second one uses a Poisson bracket identity involving the Field strength and the whole expression is thus quadratic in the field strength. Both of these regularisations do no appear in the standard case and the Hamiltonian constraint operator is thus more complicated. On the other hand, since it is already hard to deal with the usual Hamiltonian constraint, we cannot conclude that our Hamiltonian constraint is {\it significantly} more complicated. The main problem remains the simplicity constraint for which a satisfactory implementation has to be found which is compatible with the action of the Hamiltonian constraint and allows for a unitary map to the Ashtekar-Lewandowski Hilbert space.
\\
\newpage
{\bf\large Acknowledgements}\\
NB and AT thank Emanuele Alesci, Jonathan Engle, Alexander Stottmeister, and Antonia Zipfel for numerous discussions. NB and AT thank the Max Weber-Programm,  the German National Merit Foundation, and the Leonardo-Kolleg of the FAU Erlangen-N\"{u}rnbeg for financial support. NB further acknowledges financial support by the Friedrich Naumann Foundation. The part of the research performed at the Perimeter Institute for Theoretical Physics was supported in part by funds from the Government of Canada through NSERC and from the Province of Ontario through MEDT. During final improvements of this work, NB was supported by the NSF grant PHY-1205388 and the Eberly research funds of The Pennsylvania State University.

\vspace{20mm}

\begin{appendix}

\section{Alternative Regularisation of $D F^{-1} D$}

\label{app:Regularisierung}

It was suggested by Wieland \cite{WielandComplexAshtekarVariables} that one could simplify the $D$ constraints by using the classical identity
\be 
2 D_{[a} \sqrt{q} \pi_{b]IJ}(x) = -(D-1) \{ F_{abIJ}(x),  V(x, \epsilon)  \} \text{,}
\ee 
i.e. the torsion of the gravitational connection can be expressed using a Poisson bracket which will become a commutator in the quantum theory. Since the $D$ constraints appear quadratically in $\tilde{\mathcal{H}}$, this type of regularisation results in a more non-local operation of the Hamiltonian constraint. 

In order to apply the above identity, we recall from \cite{BTTII} that we can extend the covariant derivative in 
\be D^{ab}_{\overline{M}} = -\epsilon_{IJKL\overline{M}} \pi^{cIJ} \left( \pi^{(a|KN} D_c \pi^{b)L} \m_N \right) \ee 
by a Christoffel symbol acting on spatial indices on the constraint surface. Therefore, 
\be D^{ab}_{\overline{M}} = -\epsilon_{IJKL\overline{M}} \pi^{cIJ} \left( \pi^{(a|KN} q q^{b)d} D_c \pi_{d} \mbox{}^{L} \m_N \right) \ee 
and we calculate
\begin{eqnarray}
&&- \bar{d}_{(a|AB} \pi_{b)CD} \epsilon^{ABCD \overline{M}} \epsilon_{IJKL\overline{M}} \pi^{cIJ} \left( \pi^{(a|KN} q q^{b)d} D_{[c} \pi_{d]} \mbox{}^{L} \m_N \right) \nonumber\\
 &\approx& (D-3)! (D-1) \bar{K}^{\text{trace free}}_{aIJ} \bar{d}_{bKL}(F')^{aIJ,bKL}
\end{eqnarray}
with
\be
 (F')^{aIJ,bKL} = 2 \left(- E^{a|K]} E^{b[I} \eta^{J][L} + q q^{ab} \bar{\eta}^{K[I} \bar{\eta}^{J]L} \right) \text{.}
\ee
In order to have direct access to $ \bar{K}^{\text{trace free}}_{aIJ} $, we can invert $F'$ as
\be
(F{'}^{-1})_{bKL,cMN} =  \left( \frac{3}{4q} q_{bc} \bar{\eta}_{M[K} \bar{\eta}_{L]N} + \frac{1}{2} E_{b|M]} E_{c[K} \bar{\eta}_{L][N}\right)
\ee
and write
\begin{eqnarray}
\tilde{\mathcal{H}}-\mathcal{H} &\approx& \frac{1}{8 ((D-3)!)^2 (D-1)^2} \\ \nonumber
&& D{'}^{ab}_{\overline{M}} \epsilon^{ABCD \overline{M}}  \pi_{(b|AB}  (F{'}^{-1})_{a)CD,eMN} F^{eMN,fOP} (F{'}^{-1})_{fOP,(c|EF}  \pi_{d)GH} \epsilon^{EFGH \overline{N}} D{'}^{cd}_{\overline{N}}  \nonumber \\
&\approx& \frac{1}{8 ((D-3)!)^2 (D-1)^2} \nonumber \\ \nonumber
&& D{'}^{ab}_{\overline{M}} \epsilon^{ABCD \overline{M}}  \pi_{(b|AB}  \left( \frac{5}{4q} q_{a)(c} \bar{\eta}_{C[E} \bar{\eta}_{F]D} + \frac{3}{2} E_{a)|E]} E_{(c|[C}\bar{\eta}_{D][F}  \right)  \pi_{d)GH} \epsilon^{EFGH \overline{N}} D{'}^{cd}_{\overline{N}} \nonumber
\end{eqnarray}
with
\be D{'}^{ab}_{\overline{M}} = -\epsilon_{IJKL\overline{M}} \pi^{cIJ} \left( \pi^{(a|KN} q q^{b)d} D_{[c} \pi_{d]} \mbox{}^{L} \m_N \right) \text{.} \ee

We can also implement the above Poisson bracket identity without starting from the original $D$ constraints but by trying to find an easier expression for $\tilde{\mathcal{H}} -\mathcal{H}$ directly from $D_{[a} \pi_{b]IJ}$. It turns out that
\be
\tilde{\mathcal{H}} -\mathcal{H} \approx  \zeta (D_{[a} \sqrt{q} \pi_{b]IJ}) (D_{[c} \sqrt{q} \pi_{d]KL}) n^J n^L \left(q^{bd} E^{aK} E^{cI} +\frac{1}{2} q q^{a[c} q^{d]b} \bar{\eta}^{IK} \right) \text{.}
\ee

The obvious question at this point is which of the two expressions is suited better for a quantisation. Although a satisfactory answer might only be possible after studying the quantum dynamics, we see at the classical level that the second expression has a less complicated index structure due to the missing epsilon symbols. On the other hand, it contains correction terms proportional to $\bar{K}^{\text{tr}}_I$, which are absent due to the epsilon symbols in the first expression. In the formulation studied in this paper, this does not affect the theory since $\bar{K}^{\text{tr}}_I\approx 0$ on the constraint surface \cite{BTTII}. In general, this won't be true any more when coupling fermions \cite{BTTIV} or using the time normal $n^I$ as a independent field \cite{BTTVI} in other papers of this series. Although introducing additional correction terms, an independent time normal would simplify the expression since the action of a multiplication operator corresponding to $n^I$ is simpler than the regularised version of $n^In^J(\pi)$. \\
\\
\\

\end{appendix}

\bibliography{pa89pub.bbl}

\providecommand{\href}[2]{#2}\begingroup\raggedright\begin{thebibliography}{10}

\bibitem{BTTI}
N.~Bodendorfer, T.~Thiemann, and A.~Thurn, ``{New variables for classical and
  quantum gravity in all dimensions: I. Hamiltonian analysis},'' {\em Classical
  and Quantum Gravity} {\bf 30} (2013) 045001, {\tt arXiv:1105.3703 [gr-qc]}.

\bibitem{BTTII}
N.~Bodendorfer, T.~Thiemann, and A.~Thurn, ``{New variables for classical and
  quantum gravity in all dimensions: II. Lagrangian analysis},'' {\em Classical
  and Quantum Gravity} {\bf 30} (2013) 045002, {\tt arXiv:1105.3704 [gr-qc]}.

\bibitem{RovelliQuantumGravity}
C.~Rovelli, {\em {Quantum Gravity}}.
\newblock Cambridge University Press, Cambridge, 2004.

\bibitem{ThiemannModernCanonicalQuantum}
T.~Thiemann, {\em {Modern Canonical Quantum General Relativity}}.
\newblock Cambridge University Press, Cambridge, 2007.

\bibitem{EngleTheLoopQuantum}
J.~Engle, R.~Pereira, and C.~Rovelli, ``{The Loop-Quantum-Gravity Vertex
  Amplitude},'' {\em Physical Review Letters} {\bf 99} (2007) 161301, {\tt
  arXiv:0705.2388 [gr-qc]}.

\bibitem{LivineNewSpinfoamVertex}
E.~Livine and S.~Speziale, ``{New spinfoam vertex for quantum gravity},'' {\em
  Physical Review D} {\bf 76} (2007) 084028, {\tt arXiv:0705.0674 [gr-qc]}.

\bibitem{EngleFlippedSpinfoamVertex}
J.~Engle, R.~Pereira, and C.~Rovelli, ``{Flipped spinfoam vertex and loop
  gravity},'' {\em Nuclear Physics B} {\bf 798} (2008) 251--290, {\tt
  arXiv:0708.1236 [gr-qc]}.

\bibitem{EngleLoopQuantumGravity}
J.~Engle, E.~R. Livine, R.~Pereira, and C.~Rovelli, ``{LQG vertex with finite
  Immirzi parameter},'' {\em Nuclear Physics B} {\bf 799} (2008) 136--149, {\tt
  arXiv:0711.0146 [gr-qc]}.

\bibitem{FreidelANewSpin}
L.~Freidel and K.~Krasnov, ``{A new spin foam model for 4D gravity},'' {\em
  Classical and Quantum Gravity} {\bf 25} (2008) 125018, {\tt arXiv:0708.1595
  [gr-qc]}.

\bibitem{KaminskiSpinFoamsFor}
W.~Kaminski, M.~Kisielowski, and J.~Lewandowski, ``{Spin-foams for all loop
  quantum gravity},'' {\em Classical and Quantum Gravity} {\bf 27} (2010)
  095006, {\tt arXiv:0909.0939 [gr-qc]}.

\bibitem{AshtekarRepresentationsOfThe}
A.~Ashtekar and C.~J. Isham, ``{Representations of the holonomy algebras of
  gravity and non-Abelian gauge theories},'' {\em Classical and Quantum
  Gravity} {\bf 9} (1992) 1433--1468, {\tt arXiv:hep-th/9202053}.

\bibitem{AshtekarRepresentationTheoryOf}
A.~Ashtekar and J.~Lewandowski, ``{Representation Theory of Analytic Holonomy
  C* Algebras},'' in {\em Knots and Quantum Gravity} (J.~Baez, ed.), (Oxford),
  Oxford University Press1994.
\newblock {\tt arXiv:gr-qc/9311010}.

\bibitem{AshtekarDifferentialGeometryOn}
A.~Ashtekar and J.~Lewandowski, ``{Differential geometry on the space of
  connections via graphs and projective limits},'' {\em Journal of Geometry and
  Physics} {\bf 17} (1995) 191--230, {\tt arXiv:hep-th/9412073}.

\bibitem{AshtekarProjectiveTechniquesAnd}
A.~Ashtekar and J.~Lewandowski, ``{Projective techniques and functional
  integration for gauge theories},'' {\em Journal of Mathematical Physics} {\bf
  36} (1995) 2170--2191, {\tt arXiv:gr-qc/9411046}.

\bibitem{MarolfOnTheSupport}
D.~Marolf and J.~M. Mour\~{a}o, ``{On the support of the Ashtekar-Lewandowski
  measure},'' {\em Communications in Mathematical Physics} {\bf 170} (1995)
  583--605, {\tt arXiv:hep-th/9403112}.

\bibitem{AshtekarQuantizationOfDiffeomorphism}
A.~Ashtekar, J.~Lewandowski, D.~Marolf, J.~M. Mour\~{a}o, and T.~Thiemann,
  ``{Quantization of diffeomorphism invariant theories of connections with
  local degrees of freedom},'' {\em Journal of Mathematical Physics} {\bf 36}
  (1995) 6456--6493, {\tt arXiv:gr-qc/9504018}.

\bibitem{LewandowskiUniquenessOfDiffeomorphism}
J.~Lewandowski, A.~Okol\'{o}w, H.~Sahlmann, and T.~Thiemann, ``{Uniqueness of
  Diffeomorphism Invariant States on Holonomy-Flux Algebras},'' {\em
  Communications in Mathematical Physics} {\bf 267} (2006) 703--733, {\tt
  arXiv:gr-qc/0504147}.

\bibitem{FleischhackRepresentationsOfThe}
C.~Fleischhack, ``{Representations of the Weyl Algebra in Quantum Geometry},''
  {\em Communications in Mathematical Physics} {\bf 285} (2009) 67--140, {\tt
  arXiv:math-ph/0407006}.

\bibitem{RovelliSpinNetworksAnd}
C.~Rovelli and L.~Smolin, ``{Spin networks and quantum gravity},'' {\em
  Physical Review D} {\bf 52} (1995) 5743--5759, {\tt arXiv:gr-qc/9505006}.

\bibitem{BaezSpinNetworksIn}
J.~Baez, ``{Spin Networks in Gauge Theory},'' {\em Advances in Mathematics}
  {\bf 117} (1996) 253--272, {\tt arXiv:gr-qc/9411007}.

\bibitem{ThiemannTheInverseLoop}
T.~Thiemann, ``{The inverse loop transform},'' {\em Journal of Mathematical
  Physics} {\bf 39} (1998) 1236--1248, {\tt arXiv:hep-th/9601105}.

\bibitem{ThiemannQSD5}
T.~Thiemann, ``{Quantum spin dynamics (QSD) V: Quantum Gravity as the Natural
  Regulator of Matter Quantum Field Theories},'' {\em Classical and Quantum
  Gravity} {\bf 15} (1998) 1281--1314, {\tt arXiv:gr-qc/9705019}.

\bibitem{FreidelBFDescriptionOf}
L.~Freidel, K.~Krasnov, and R.~Puzio, ``{BF description of higher-dimensional
  gravity theories},'' {\em Advances in Theoretical and Mathematical Physics}
  {\bf 3} (1999) 1289--1324, {\tt arXiv:hep-th/9901069}.

\bibitem{SahlmannIrreducibilityOfThe}
H.~Sahlmann and T.~Thiemann, ``{Irreducibility of the
  Ashtekar-Isham-Lewandowski representation},'' {\em Classical and Quantum
  Gravity} {\bf 23} (2006) 4453--4471, {\tt arXiv:gr-qc/0303074}.

\bibitem{BTTV}
N.~Bodendorfer, T.~Thiemann, and A.~Thurn, ``{On the implementation of the
  canonical quantum simplicity constraint},'' {\em Classical and Quantum
  Gravity} {\bf 30} (2013) 045005, {\tt arXiv:1105.3708 [gr-qc]}.

\bibitem{ThiemannThePhoenixProject}
T.~Thiemann, ``{The Phoenix Project: master constraint programme for loop
  quantum gravity},'' {\em Classical and Quantum Gravity} {\bf 23} (2006)
  2211--2247, {\tt arXiv:gr-qc/0305080}.

\bibitem{ThiemannClosedFormulaFor}
T.~Thiemann, ``{Closed formula for the matrix elements of the volume operator
  in canonical quantum gravity},'' {\em Journal of Mathematical Physics} {\bf
  39} (1998) 3347--3371, {\tt arXiv:gr-qc/9606091}.

\bibitem{ThiemannQSD3}
T.~Thiemann, ``{Quantum spin dynamics (QSD) III: Quantum constraint algebra and
  physical scalar product in quantum general relativity},'' {\em Classical and
  Quantum Gravity} {\bf 15} (1998) 1207--1247, {\tt arXiv:gr-qc/9705017}.

\bibitem{SmolinRecentDevelopmentsIn}
L.~Smolin, ``{Recent developments in non-perturbative quantum gravity},'' {\tt
  arXiv:hep-th/9202022}.

\bibitem{AshtekarQuantumTheoryOf1}
A.~Ashtekar and J.~Lewandowski, ``{Quantum theory of gravity I: Area
  operators},'' {\em Classical and Quantum Gravity} {\bf 14} (1997) A55--A81,
  {\tt arXiv:gr-qc/9602046}.

\bibitem{GirardiGeneralizedYoungTableaux}
G.~Girardi, A.~Sciarrino, and P.~Sorba, ``{Generalized Young tableaux and
  Kronecker products of SO(n) representations},'' {\em Physica A: Statistical
  Mechanics and its Applications} {\bf 114} (1982) 365--369.

\bibitem{GirardiKroneckerProductsFor}
G.~Girardi, A.~Sciarrino, and P.~Sorba, ``{Kronecker products for SO(2p)
  representations},'' {\em Journal of Physics A: Mathematical and General} {\bf
  15} (1982) 1119--1129.

\bibitem{mourao1999}
J.~Mour\~{a}o, T.~Thiemann, and J.~Velhinho, ``{Physical properties of quantum
  field theory measures},'' {\em Journal of Mathematical Physics} {\bf 40}
  (1999) 2337--2353, {\tt arXiv:hep-th/9711139}.

\bibitem{GieselConsistencyCheckOn1}
K.~Giesel and T.~Thiemann, ``{Consistency check on volume and triad operator
  quantization in loop quantum gravity: I},'' {\em Classical and Quantum
  Gravity} {\bf 23} (2006) 5667--5691, {\tt arXiv:gr-qc/0507036}.

\bibitem{GieselConsistencyCheckOn2}
K.~Giesel and T.~Thiemann, ``{Consistency check on volume and triad operator
  quantization in loop quantum gravity: II},'' {\em Classical and Quantum
  Gravity} {\bf 23} (2006) 5693--5771, {\tt arXiv:gr-qc/0507037}.

\bibitem{BrunnemannSimplificationOfThe}
J.~Brunnemann and T.~Thiemann, ``{Simplification of the Spectral Analysis of
  the Volume Operator in Loop Quantum Gravity},'' {\em Classical and Quantum
  Gravity} {\bf 23} (2006) 1289--1346, {\tt arXiv:gr-qc/0405060}.

\bibitem{BrunnemannPropertiesOfThe1}
J.~Brunnemann and D.~Rideout, ``{Properties of the volume operator in loop
  quantum gravity: I. Results},'' {\em Classical and Quantum Gravity} {\bf 25}
  (2008) 065001, {\tt arXiv:0706.0469 [gr-qc]}.

\bibitem{BrunnemannPropertiesOfThe2}
J.~Brunnemann and D.~Rideout, ``{Properties of the volume operator in loop
  quantum gravity: II. Detailed presentation},'' {\em Classical and Quantum
  Gravity} {\bf 25} (2008) 065002, {\tt arXiv:0706.0382 [gr-qc]}.

\bibitem{BrunnemannOrientedMatroidsCombinatorial}
J.~Brunnemann and D.~Rideout, ``{Oriented matroids - combinatorial structures
  underlying loop quantum gravity},'' {\em Classical and Quantum Gravity} {\bf
  27} (2010) 205008, {\tt arXiv:1003.2348 [gr-qc]}.

\bibitem{ThiemannQSD1}
T.~Thiemann, ``{Quantum spin dynamics (QSD)},'' {\em Classical and Quantum
  Gravity} {\bf 15} (1998) 839--873, {\tt arXiv:gr-qc/9606089}.

\bibitem{ThiemannQSD2}
T.~Thiemann, ``{Quantum spin dynamics (QSD): II. The kernel of the
  Wheeler-DeWitt constraint operator},'' {\em Classical and Quantum Gravity}
  {\bf 15} (1998) 875--905, {\tt arXiv:gr-qc/9606090}.

\bibitem{ThiemannQSD8}
T.~Thiemann, ``{Quantum spin dynamics: VIII. The master constraint},'' {\em
  Classical and Quantum Gravity} {\bf 23} (2006) 2249--2265, {\tt
  arXiv:gr-qc/0510011}.

\bibitem{ThiemannQSD4}
T.~Thiemann, ``{QSD 4: (2+1) Euclidean quantum gravity as a model to test (3+1)
  Lorentzian quantum gravity},'' {\em Class.Quant.Grav.} {\bf 15} (1998)
  1249--1280, {\tt arXiv:gr-qc/9705018}.

\bibitem{DingTheVolumeOperator}
Y.~Ding and C.~Rovelli, ``{The volume operator in covariant quantum gravity},''
  {\tt arXiv:0911.0543 [gr-qc]}.

\bibitem{BTTIV}
N.~Bodendorfer, T.~Thiemann, and A.~Thurn, ``{New variables for classical and
  quantum gravity in all dimensions: IV. Matter coupling},'' {\em Classical and
  Quantum Gravity} {\bf 30} (2013) 045004, {\tt arXiv:1105.3706 [gr-qc]}.

\bibitem{BTTVI}
N.~Bodendorfer, T.~Thiemann, and A.~Thurn, ``{Towards loop quantum supergravity
  (LQSG): I. Rarita-Schwinger sector},'' {\em Classical and Quantum Gravity}
  {\bf 30} (2013) 045006, {\tt arXiv:1105.3709 [gr-qc]}.

\bibitem{BTTVII}
N.~Bodendorfer, T.~Thiemann, and A.~Thurn, ``{Towards loop quantum supergravity
  (LQSG): II. p -form sector},'' {\em Classical and Quantum Gravity} {\bf 30}
  (2013) 045007, {\tt arXiv:1105.3710 [gr-qc]}.

\bibitem{WielandComplexAshtekarVariables}
W.~Wieland, ``{Complex Ashtekar variables and reality conditions for Holst's
  action},'' {\em Annales Henri Poincar\'{e}} {\bf 13} (2012), no.~3 425--448,
  {\tt arXiv:1012.1738 [gr-qc]}.

\end{thebibliography}\endgroup

\end{document}